\documentclass[preprint,12pt]{elsarticle}

\usepackage[vlined]{algorithm2e}
\usepackage{url}
\usepackage{tikz} 
\usetikzlibrary{arrows.meta}

\usepackage{amsmath,amssymb,amsfonts}
\usepackage{amsthm,amscd,bm}
\usepackage[a4paper,top=3cm,bottom=3cm,left=3cm,right=3cm]{geometry}
\usepackage{makecell}

\newcommand{\N}{\ensuremath{\mathbb{N}}\xspace}

\newcommand{\Fp}{\mathbb{F}_p}
\newcommand{\KK}{\mathbb{K}}
\newcommand{\SSS}{\mathcal{S}}

\newcommand{\myvec}[1]{\ensuremath{\boldsymbol{#1}}\xspace}

\newcommand{\entropy}{\mathcal{H}}
\newcommand{\RWT}{\mathcal{R}}
\newcommand{\cyl}[1]{{Cyl(#1)}}

\newcommand{\x}{\myvec{x}}

\newcommand{\ZZ}{\mathbb{Z}}
\newcommand{\zzd}{{\mathbb{Z}^D}}
\newcommand{\LP}{\mathbb{L}_p}

\newcommand{\Aut}{\mathrm{Aut}}
\newcommand{\Ker}{\mathrm{Ker}}
\newcommand{\End}{\mathrm{End}}

\newcommand{\Imma}{\mathrm{Im}}

\newcommand{\locrule}{\ensuremath{f}}
\newcommand{\glorule}{\ensuremath{{\cal F}}}

\usepackage{xspace, comment, mathdots}

\newcommand{\ie}{i.e.\@\xspace}
\newcommand{\eg}{e.g.\@\xspace}

\DeclareMathOperator{\lcm}{lcm}
\DeclareMathOperator{\rk}{rk}
\DeclareMathOperator{\Supp}{Supp}
\DeclareMathOperator{\supp}{supp}
\DeclareMathOperator{\wdt}{wd}

\newcommand{\opdelta}{\ensuremath{{|\delta|}}}

\newcommand{\lw}{\mathit{T^-}}
\newcommand{\rw}{\mathit{T^+}}
\newcommand{\w}{\mathit{T}}
\newcommand{\tr}{\mathrm{tr}}

\newtheorem{theorem}{Theorem}
\newtheorem{definition}{Definition}
\newtheorem{remark}{Remark}

\newtheorem{lemma}{Lemma}

\newtheorem{corollary}{Corollary}

\newtheorem{proposition}{Proposition}

\begin{document}

\begin{frontmatter}

\title{Topological Entropy of Group\\ Cellular Automata is Effectively Computable}
\author[unibo]{Niccol\`o Castronuovo}
\ead{niccolo.castronuovo2@unibo.it}

\author[unimib]{Alberto Dennunzio\corref{cor}}
\ead{alberto.dennunzio@unimib.it}

\author[UTU]{Jarkko Kari}
\ead{jkari@utu.fi}


\author[unibo]{Luciano Margara}
\ead{luciano.margara@unibo.it}
\cortext[cor]{Corresponding author}

\affiliation[unimib]{organization={Dipartimento di Informatica, Sistemistica e Comunicazione,
  Universit\`a degli Studi di Milano-Bicocca},
            addressline={Viale Sarca 336/14}, 
            city={Milano},
            postcode={20126}, 
            country={Italy}}
            
  
\address[UTU]{Department of Mathematics and Statistics, University of Turku, Finland}


\affiliation[unibo]{organization={Department of Computer Science and Engineering, University of Bologna, Cesena Campus},
            addressline={Via dell'Universit\`a 50}, 
            city={Cesena},
            postcode={47521}, 
            country={Italy}}
            
\begin{abstract}
Topological entropy is a fundamental numerical invariant of dynamical systems   capturing in a quantitative way their dynamical complexity.
Determining the topological entropy of cellular automata is a notoriously
difficult problem and is, in general, algorithmically intractable. 
In this paper we study \emph{group cellular automata}, namely cellular automata over a finite group alphabet whose global evolution map is a group endomorphism.

In dimension one, our main contribution is a structural decomposition of the topological entropy 
into two one-sided components that capture, separately, the information propagating from the left and from the right in the space--time evolution. 
We introduce finite families of left and right space--time 
traces and show that the topological entropy of a group cellular automaton can be expressed as the
sum of the associated left and right growth rates.
As a consequence, the topological entropy is completely determined by the growth
rate of a fixed-width family of space--time patterns arising in the evolution of
the automaton. This reduction from arbitrarily large space--time windows to a
uniformly bounded width yields an effective procedure for computing the
topological entropy. As a by-product, the topological entropy of a one-dimensional group cellular automaton is always the logarithm of a positive integer.

Using Pontryagin duality and the Bridge Theorem relating topological and algebraic entropy, we also give an explicit formula for the topological entropy of linear cellular automata over $(\mathbb{Z}/p\mathbb{Z})^n$ in terms of their characteristic polynomial, and we prove that, in every dimension $D\geq 2$, a group cellular automaton is either equicontinuous with zero topological entropy, or sensitive to initial conditions with infinite topological entropy. Together with the decidability of equicontinuity, this shows that the topological entropy of group cellular automata is effectively computable from the local rule, in every dimension.

These results single out group cellular automata as a natural and nontrivial
class in which topological entropy becomes computable, revealing the role of
algebraic constraints in restoring algorithmic tractability.
\end{abstract}



\end{frontmatter}



\section{Introduction} \label{introduction}

Cellular automata (CAs) are discrete-time dynamical systems defined by the synchronous
iteration of a local rule on an infinite lattice.
They were originally introduced by von Neumann as abstract models of
self-reproducing systems~\cite{vonNeumann1966}, and later studied systematically
within symbolic dynamics and theoretical computer science.
A cellular automaton consists of a regular grid of variables, typically indexed by
$\mathbb{Z}^D$, each taking values in a finite alphabet, and evolving in discrete
time according to a rule that updates every variable based only on the states of a
finite neighborhood (for an introduction to CA theory, see, e.g.,~\cite{CeccheriniSilbersteinCoornart2010, Hadeler2017CellularAA, Hedlund1969}).

Despite the simplicity of their local definition, cellular automata are capable
of generating a wide range of dynamical behaviors, from highly regular dynamics
to complex and chaotic evolutions (see, e.g.,~\cite{Wolfram1983}).

CAs have been the subject of significant research and are applied in various fields, including computer science, physics, mathematics, biology, and chemistry, for purposes such as simulating natural phenomena, generating pseudo-random numbers, processing images, analyzing universal computation models, and cryptography (see for example~\cite{AnghelescuIS07,MARTINDELREY20051356,DennunzioFGM21INS,kari2000,RubioEWRS04}).

From a mathematical perspective, the Curtis-Hedlund-Lyndon theorem provides a
complete characterization of cellular automata as exactly the continuous maps
on the full shift that commute with the shift action~\cite{Hedlund1969}.

A central goal in the theory of cellular automata is to quantify dynamical
complexity.
Among the most important invariants, \emph{topological entropy} measures the
asymptotic rate at which distinguishable orbit segments proliferate, and it is
widely used as an indicator of chaotic behavior in symbolic dynamics and
cellular systems (see for example~\cite{Blanchard1992,Hurd_Kari_Culik_1992,LindMarcus1995,Shereshevsky1991}). 
For cellular automata, entropy admits a concrete combinatorial
interpretation: it can be expressed through the growth rate of the number of
distinct space-time patterns occurring in the evolution of the automaton~\cite{Hurd_Kari_Culik_1992} (see Proposition~\ref{prop:Ddim} for arbitrary dimension).
This viewpoint makes entropy particularly natural for CAs, as it directly
quantifies the production of new patterns in space-time.

The problem of computing topological entropy for cellular automata has been
extensively studied and is known to exhibit strong algorithmic limitations.
In a seminal paper, Hurd, Kari, and Culik proved that the topological
entropy of one-dimensional cellular automata is not computable, showing in
particular that there is no algorithm deciding whether the entropy is zero~\cite{Hurd_Kari_Culik_1992}.
This result established topological entropy as an intrinsically intractable
invariant for general cellular automata and initiated a line of research on its
computational complexity.

Subsequent work further clarified the scope of these negative results.
Guillon and Zinoviadis showed that the entropies of one-dimensional cellular
automata are exactly the right-computable real numbers, while in dimension two
the class of achievable entropies becomes strictly larger, encompassing
$\Sigma_2$ real numbers and infinity~\cite{GuillonZinoviadis2012}.
These results provide a precise computability-theoretic classification of
cellular automaton entropies, highlighting a sharp distinction between
dimensions.
More recently, it has been shown that undecidability persists even under strong
dynamical constraints, such as reversibility, as the problem of deciding whether
a reversible one-dimensional cellular automaton has zero entropy remains
undecidable~\cite{Hotanen2023ReversibleEntropy}.

On the positive side, computability of entropy has been established for several
restricted but structurally meaningful classes of cellular automata.
D'Amico, Manzini, and Margara proved that the entropy is computable for linear
cellular automata over $\ZZ/m\ZZ$ and for positively expansive cellular
automata~\cite{DamicoMM03}.
These results indicate that additional algebraic or dynamical rigidity can
restore algorithmic tractability.

The present work fits into this latter direction.
By exploiting the algebraic structure of a finite group alphabet and the
endomorphism property of the global rule, we identify a new, natural class of
one-dimensional cellular automata for which topological entropy becomes
effectively computable.
Our results also settle questions left open in the literature. In~\cite{CASTRONUOVO2026103749} the computation of the topological entropy of a GCA was reduced to the case of alphabets that are either elementary abelian groups $(\ZZ/p\ZZ)^n$ or products of isomorphic non-abelian simple groups, and was carried out only in the surjective non-abelian case; the abelian case is treated here (Theorem~\ref{formula}), while the non-surjective non-abelian case is avoided by a refinement of the decomposition, based on normal $\glorule$-invariant subgroups instead of fully invariant ones (Lemma~\ref{lem:terminal}).
In higher dimensions, Shereshevsky~\cite{Shereshevsky1993Indag} showed that a cellular automaton on the two-dimensional full shift cannot be expansive, and conjectured that its topological entropy is either $0$ or $+\infty$ (see~\cite{MorrisWard1998}). For general CAs the conjecture does not hold, since in every dimension $D\geq2$ there exist CAs with finite positive entropy~\cite{Meyerovitch2008}. Morris and Ward pointed out that a special case of the conjecture concerns CAs which are endomorphisms of the group structure of the full shift; they formulated an algebraic analogue of the question and answered it for mixing endomorphisms of compact abelian groups commuting with a completely positive entropy $\ZZ^d$-action~\cite{MorrisWard1998}. Theorem~\ref{thm:dichotomy} settles this special case in every dimension $D\geq2$, for every finite group alphabet and without any mixing assumption: the topological entropy of a $D$-GCA is either $0$ or $+\infty$.


In this paper we focus on \emph{group cellular automata} (GCAs).
Let $G$ be a finite group.
The full shift $G^{\mathbb Z}$ inherits a natural group structure via
component-wise multiplication, and a GCA is a CA whose global map
$\mathcal F:G^{\mathbb Z}\to G^{\mathbb Z}$ is a group endomorphism.
Such systems arise naturally in the study of algebraic dynamics and have been
systematically investigated in the literature~\cite{BeaurK24,CASTRONUOVO2026103749,
GCA24}.

Our main result shows that the algebraic constraints imposed by the group
structure lead to a strong structural simplification of the entropy computation.
We introduce two families of finite objects that capture, separately, the
left-moving and right-moving information in space-time.
More precisely, for each time $t$ we consider the sets of left and right traces
$\lw(\mathcal F,t)$ and $\rw(\mathcal F,t)$ obtained by evolving configurations
that are trivial on one side.
We then define two one-sided growth rates
\[
\entropy^{-}(\mathcal F)=\lim_{t\to\infty}\frac{\log|\lw(\mathcal F,t)|}{t},
\qquad
\entropy^{+}(\mathcal F)=\lim_{t\to\infty}\frac{\log|\rw(\mathcal F,t)|}{t},
\]
and we prove (Theorems~\ref{wall1} and~\ref{hat}) that for every one-dimensional group cellular automaton $\mathcal F$ on a
finite group alphabet $G$, the topological entropy satisfies
\[
\entropy(\mathcal F)=\entropy^{-}(\mathcal F)+\entropy^{+}(\mathcal F).
\]

\medskip
As a consequence, the topological entropy is completely determined by the growth
rate of a fixed width family of space-time patterns, denoted by  $\w(\glorule,t,0)$,
arising in the evolution of
the automaton. This reduction from arbitrarily large space-time windows to a
uniformly bounded width yields an effective procedure for computing the
topological entropy (Theorem~\ref{computable}).
We also show that the topological entropy of a one-dimensional GCA is always the logarithm of a positive integer dividing $|G|^{l_\glorule+r_\glorule}$, where $l_\glorule$ and $r_\glorule$ are the left and right radii of $\glorule$ (Corollary~\ref{cor:logint}).

In Section~\ref{sec:linear} we turn to linear GCAs over $(\ZZ/p\ZZ)^n$, whose global rule is described by a matrix $M$ of Laurent polynomials. Following the algebraic approach of Ak\i n, Dikranjan, Giordano Bruno and Toller~\cite{ADGT2024}, we pass to the Pontryagin dual, on which the dual of the automaton acts on the discrete group of finite configurations, and we use the Bridge Theorem~\cite{DikranjanGiordanoBruno2012,Weiss1974} relating topological and algebraic entropy. The algebraic entropy turns out to be the rank of a module over the polynomial ring $\Fp[t]$, which can be read off the characteristic polynomial $\chi_M$ of $M$. In dimension one this yields
\[
\entropy(\glorule)=(E^+_M-E^-_M)\log p,
\]
where $[E^-_M,E^+_M]$ is the smallest interval containing $0$ and all exponents appearing in the coefficients of $\chi_M$ (Theorem~\ref{formula}); in dimension $D\geq2$ the entropy is either $0$ or $+\infty$. Combined with a refinement of the decomposition of GCAs introduced in~\cite{CASTRONUOVO2026103749}, this provides a second, explicit, proof of the computability of the topological entropy of one-dimensional GCAs (Remark~\ref{rem:secondproof}).
Finally, we prove that a $D$-dimensional GCA, with $D\geq 2$, is either equicontinuous with zero topological entropy, or sensitive to initial conditions with infinite topological entropy (Theorem~\ref{thm:dichotomy}).
As a consequence, the topological entropy of group cellular automata is computable in every dimension (Corollary~\ref{cor:computableD}).
Finally, we give an explicit algorithm, running in polynomial time for every fixed dimension, which computes the topological entropy of a GCA from its local rule (Theorem~\ref{thm:complexity}).


The paper is organized as follows.
Section~\ref{CAandGCA} recalls the basic definitions of CAs and GCAs in arbitrary dimension.
In Section~\ref{sec:entropy} we review topological entropy and its space-time characterization
for cellular automata.
Section~\ref{sec:mainresult} contains the proof of the decomposition theorem and
the resulting computability statement.
Section~\ref{sec:linear} contains the explicit formula for linear GCAs over $(\ZZ/p\ZZ)^n$ in every dimension.
Section~\ref{sec:Ddim} deals with $D$-dimensional GCAs, $D\geq 2$.
Section~\ref{sec:complexity} analyzes the computational complexity of the resulting procedure.

\section{CAs and GCAs} \label{CAandGCA}

In this section, we review the fundamental definitions and key results related to CAs and GCAs. For additional definitions and results, we refer the reader to those introduced in~\cite{GCA24}.
We give the definitions in an arbitrary dimension $D\geq1$; at the beginning of each of the following sections we specify whether the dimension is arbitrary or $D=1$.

\medskip\noindent
\textbf{Configurations.}
Let $D\geq1$ be an integer and let $G$ be a finite set. A ($D$-dimensional) \emph{configuration} is any function from $\zzd$ to $G$, \ie, an element of $G^\zzd$. Given a configuration $c\in G^\zzd$ and $v\in\zzd$, the value of $c$ at position $v$ is denoted by $c_v$, and for $B\subseteq\zzd$ we denote by $c|_B$ the restriction of $c$ to $B$. For $v\in\zzd$ we let $\|v\|_\infty=\max_i|v_i|$, and for $w\geq0$ we let $B_w=[-w,w]^D$.
We equip $G^\zzd$ with the prodiscrete topology, that is, the product
topology obtained by endowing each factor $G$ with the discrete topology.
With this topology, $G^\zzd$ is a compact topological space. Since $G$ is finite, $G^\zzd$ is also metrizable.
Indeed, the set $G^\zzd$ can be equipped with the standard Tychonoff distance $d$ 
defined as 
$$
\forall c,c^\prime\in G^\zzd:\  d(c,c')=\begin{cases}
0 & \text{ if }  c=c' \\
2^{-\Delta(c,c')} &\text{otherwise}
\end{cases}
$$
where $\Delta(c,c')=\min\{\|j\|_\infty: j\in\zzd \text{ and } c_j\neq c'_j\}.$ 
The topology induced by the Tychonoff distance coincides with the prodiscrete topology. Note that $d(c,c')<2^{-w}$ if and only if $c$ and $c'$ coincide on $B_w$.
For a finite set $B\subseteq\zzd$ and $u\in G^B$, the \emph{cylinder} determined by $u$ is the set
\[
\cyl{B,u}=\{c\in G^\zzd:\ c|_B=u\}.
\]
Cylinders are clopen and form a basis of the prodiscrete topology; in particular, for every $c\in G^\zzd$ and $w\geq0$, the open ball of radius $2^{-w}$ centered at $c$ is the cylinder $\cyl{B_w,c|_{B_w}}$.

\medskip\noindent
\textbf{Cellular automata.}
For $w\in\zzd$, the \emph{shift} $\sigma^w:G^\zzd\to G^\zzd$ is defined by
$$\forall c\in G^\zzd,\ \forall v\in\zzd:\ \sigma^w(c)_v=c_{v+w}.$$
A ($D$-dimensional) \emph{CA} on $G$ is any continuous function $\glorule: G^\zzd \to G^\zzd$ which is also shift commuting, \ie, $\glorule\circ \sigma^w=\sigma^w \circ \glorule$ for every $w\in\zzd$.
By the Curtis-Hedlund-Lyndon theorem~\cite{Hedlund1969}, any CA can be equivalently defined by means of a \emph{local rule} $f:G^k \to G$ together with a finite set $V_f=\{v_1,\dots,v_k\}\subseteq\zzd$ of pairwise distinct vectors, called the \emph{neighborhood}, as follows:
$$\forall c \in G^\zzd,\ \forall v\in \zzd:\ \glorule(c)_v=f(c_{v+v_1},\dots, c_{v+v_k}).$$

\medskip\noindent
\textbf{Group cellular automata.}
Let $G$ be a finite group with identity element $e$. The set $G^\zzd$ is
itself a group under the component-wise operation induced by the group operation
of $G$. We denote by $e^\zzd \in G^\zzd$ the configuration taking
the value $e$ at every position; that is, $e^\zzd$ is the
identity element of the group $G^\zzd$. Clearly, when equipped with the prodiscrete topology, $G^\zzd$ turns out to be a compact topological group. A configuration $c\in G^\zzd$ is said to be \emph{finite} if the number of positions $v\in\zzd$ such that $c_v\neq e$ is finite.
A ($D$-dimensional) CA $\glorule: G^\zzd\to G^\zzd$ is said to be a \emph{$D$-GCA} if $\glorule$ is an endomorphism of the group $G^\zzd$. In that case, the local rule of $\glorule$ is a homomorphism $f:G^k \to G$.
For $1\leq i\leq k$, setting $h_i(g)=f(e,\dots,e,g,e,\dots,e)$, with $g$ in the $i$-th position, defines endomorphisms $h_1,\dots,h_k$ of $G$ with pairwise commuting images, and $f(g_1,\dots,g_k)=h_1(g_1)\cdots h_k(g_k)$ for all $g_1,\dots,g_k\in G$; conversely, every family of endomorphisms with pairwise commuting images defines in this way a homomorphism $G^k\to G$ (see~\cite[Thm.~1]{GCA24}). We then write $f=(h_1,\dots,h_k)$.

\medskip\noindent
\textbf{The one-dimensional case.}
When $D=1$, or when the dimension is clear from the context, we simply speak of CAs and GCAs. For $D=1$ we write $\sigma=\sigma^{1}$, so that $\sigma(c)_i=c_{i+1}$, and we denote by $\rho=\sigma^{-1}$ the right shift, \ie, $\rho(c)_i=c_{i-1}$. For $i\leq j$ we write $c_{[i,j]}=c_i\cdots c_j\in G^{j-i+1}$. Moreover, for a one-dimensional CA $\glorule$ with neighborhood $V_f\subseteq\ZZ$ we define $l_\glorule$ and $r_\glorule$ as follows
\begin{equation}\label{landr}
    l_\glorule= |\min(V_f\cup \{0\})| \text{  and  } r_\glorule=|\max(V_f\cup \{0\})|,
\end{equation}
so that $\glorule(c)_i$ depends only on $c_{[i-l_\glorule,\,i+r_\glorule]}$.

\section{Topological entropy}\label{sec:entropy}

\noindent In this section the dimension $D\geq1$ is arbitrary, unless otherwise stated.

A central problem in the qualitative theory of dynamical systems is to quantify
the complexity of the long-term behavior of orbits in a way that is intrinsic
and robust under natural equivalence relations.
Topological entropy was introduced precisely to address this issue, providing a
numerical invariant that measures the asymptotic complexity of a dynamical
system without relying on any probabilistic structure~\cite{AdlerKonheimMcAndrew1965}.
At an intuitive level, topological entropy measures the exponential growth rate
of the number of orbit segments that can be distinguished as the system evolves
over time. Two orbit segments are regarded as distinguishable if they eventually
separate by more than a prescribed resolution, and positive entropy reflects
the fact that the number of such distinguishable behaviors grows exponentially
when the observation time increases. In this sense, topological entropy provides
a quantitative measure of the intrinsic complexity of the dynamics, capturing
how rapidly new dynamical patterns are generated.
In systems with zero topological entropy, orbit segments remain highly
organized, and the amount of information needed to describe the dynamics grows
at most sub-exponentially.
In contrast, positive topological entropy signals the presence of complex
behavior, characterized by an exponential proliferation of distinguishable
orbit segments.
For this reason, topological entropy is widely regarded as a fundamental
indicator of deterministic chaos~\cite{Bowen1971}.
One of the key features that makes topological entropy particularly powerful is
its invariance under topological conjugacy.
As a consequence, it provides a robust tool for comparing and classifying
dynamical systems up to topological equivalence.
Moreover, topological entropy plays a unifying role within dynamical systems
theory.
Through the variational principle, it is related to measure-theoretic entropy
by being equal to the supremum of metric entropies over all invariant
probability measures~\cite{Goodman1971,Walters1982}.
This connection bridges the topological and ergodic descriptions of dynamical
behavior.
Since its introduction, topological entropy has become one of the most fundamental and widely
used invariants in the study of dynamical systems, with applications ranging
from symbolic dynamics and CA to smooth and geometric dynamics.

\medskip\noindent
There are two main and standard approaches to the definition of topological
entropy for dynamical systems.
The first one, introduced by Adler, Konheim, and McAndrew, is purely topological
and is based on open covers of the phase space~\cite{AdlerKonheimMcAndrew1965}.
The second one, developed independently by Dinaburg and Bowen, is metric in
nature and relies on the notion of separation of orbits with respect to a
compatible distance~\cite{Bowen1971,Dinaburg1970}.
The equivalence between the open cover and metric definitions of topological
entropy is a classical result.

On compact metrizable spaces the topological entropy defined via open covers in the sense of
Adler-Konheim-McAndrew coincides with the topological entropy defined via
\((n,\varepsilon)\)-separated (or spanning) sets associated with any compatible
metric (see for example ~\cite[Theorem~7.8]{Walters1982} and~\cite[Section~3.1]{katokhasselblatt1995}).

The argument relies crucially on compactness, which allows one to relate finite
open covers to metric balls and to control the growth of dynamical refinements.
As a consequence, topological entropy is a topological invariant and does not
depend on the particular compatible metric chosen.
This equivalence justifies the interchangeable use of the two approaches: the
open cover definition provides an intrinsic, purely topological formulation,
while the metric definition is often more convenient for explicit computations
and applications.

\medskip
\noindent
We now give the formal definition of topological entropy in the sense of Bowen and Dinaburg. Although in the one-dimensional part of this paper (Section~\ref{sec:mainresult}) we only use the equivalent combinatorial formulation for cellular automata given below, the general definition is needed in Sections~\ref{sec:linear} and~\ref{sec:Ddim}, where we use the Bridge Theorem and general properties of topological entropy.

\medskip
\noindent
\textbf{The Bowen-Dinaburg definition of topological entropy}. 
Assume that \(X\) is a compact metrizable space, let \(d\) be a metric
compatible with its topology, and let $T:X\to X$ be continuous.
We define the
\emph{dynamical metric}
\[
d_n(x,y)
\;=\;
\max_{0\le k<n} d(T^k x, T^k y).
\]
Given \(\varepsilon>0\), a set \(E\subset X\) is said to be
\emph{\((n,\varepsilon)\)-separated} if
\[
x\neq y \in E
\quad \Rightarrow \quad
d_n(x,y)\ge \varepsilon.
\]
Let \(s(n,\varepsilon)\) denote the maximal cardinality of an
\((n,\varepsilon)\)-separated set. By compactness, \(s(n,\varepsilon)<+\infty.\)
The topological entropy of \(T\) is defined as
\[
\entropy(T)
\;=\;
\lim_{\varepsilon\to 0}
\limsup_{n\to\infty}
\frac{1}{n}\log s(n,\varepsilon),
\]
as originally introduced in~\cite{Bowen1971} (see also~\cite[Chapter~7]{Walters1982}).
Equivalent formulations can be given using \((n,\varepsilon)\)-spanning sets
instead of separated sets.

Intuitively, two points are considered indistinguishable up to time \(n\) if their orbit
segments remain within distance \(\varepsilon\) at all intermediate times.
The quantity \(s(n,\varepsilon)\) counts the maximum number of orbit segments
that can be distinguished with resolution \(\varepsilon\) over a time window
of length \(n\).
The entropy is the asymptotic exponential growth rate of this number as
\(n\to\infty\), observed at arbitrarily fine scales.

\medskip\noindent
\textbf{Some standard facts.}
Let $T:X\to X$ and $S:Y\to Y$ be continuous maps of compact metrizable spaces. We shall use the following well-known properties, where entropies may take the value $+\infty$ (see~\cite[Chapter~7 and Chapter~8]{Walters1982}):
(a) $\entropy(T^N)=N\,\entropy(T)$ for every $N\geq 1$;
(b) if $\pi:X\to Y$ is a continuous surjection with $\pi\circ T=S\circ\pi$, then $\entropy(S)\leq\entropy(T)$;
(c) if $Z\subseteq X$ is closed and $T(Z)\subseteq Z$, then $\entropy(T|_Z)\leq\entropy(T)$; moreover $\entropy(T)=\entropy(T|_{T(X)})$ \cite[Corollary~8.6.1(iii)]{Walters1982}; 
(d) $\entropy(T\times S)=\entropy(T)+\entropy(S)$.

Recall that a continuous map $T:X\to X$ of a metric space is \emph{equicontinuous} if for every $\varepsilon>0$ there is $\eta>0$ such that $d(x,y)<\eta$ implies $d(T^kx,T^ky)<\varepsilon$ for every $k\geq0$, and that $T$ is \emph{sensitive to initial conditions} if there is $\varepsilon>0$ such that for every $x\in X$ and every $\eta>0$ there exist $y\in X$ and $k\geq0$ with $d(x,y)<\eta$ and $d(T^kx,T^ky)\geq\varepsilon$. The two properties are mutually exclusive on non-empty spaces.

\begin{proposition}\label{prop:equi}
Let $T:X\to X$ be a continuous map of a compact metric space. If $T$ is equicontinuous then $\entropy(T)=0$.
\end{proposition}
\begin{proof}
Fix $\varepsilon>0$. By equicontinuity there exists $\eta>0$ such that $d(x,y)<\eta$ implies $d(T^kx,T^ky)<\varepsilon$ for every $k\geq 0$, \ie, $d_n(x,y)<\varepsilon$ for every $n$. Hence the points of an $(n,\varepsilon)$-separated set are pairwise at distance at least $\eta$, and $s(n,\varepsilon)\leq s(1,\eta)$ for every $n$. By compactness $s(1,\eta)<\infty.$ Therefore $\limsup_n \frac1n\log s(n,\varepsilon)=0$ for every $\varepsilon$.
\end{proof}

\begin{proposition}\label{prop:GCAequi}
Let $\glorule$ be a $D$-dimensional CA on a finite alphabet $A$, $D\geq1$.
\begin{itemize}
\item[(i)] $\glorule$ is equicontinuous if and only if there exist $0\leq a<b$ such that $\glorule^a=\glorule^b$.
\item[(ii)] If $\glorule$ is a $D$-GCA, then $\glorule$ is either equicontinuous or sensitive to initial conditions.
\end{itemize}
\end{proposition}
\noindent Part (i) is~\cite[Proposition~84]{KariNotes}, and part (ii) is~\cite[Lemma~26]{BeaurK24}, where it is proved for group cellular automata on arbitrary group shifts (see also the proof of~\cite[Theorem~27]{BeaurK24} for (i) in the case of GCAs).

\medskip
\noindent 
\textbf{Topological entropy for CA.} 
In~\cite{Hurd_Kari_Culik_1992} it is shown that for one-dimensional CA
the general definition of topological entropy translates to the following simpler form.
For $D=1$, let $\RWT(w,t)$ denote the number of distinct rectangles of width $w$ and height $t+1$ occurring in a space-time evolution diagram of $\glorule$ (see Figure~\ref{fig:trace} and Equation~\eqref{R as trace} below for a formal definition). 
Then, 
\begin{equation}\label{entropydef}
    \entropy (\glorule)=\lim_{w\to +\infty}\left( \lim_{t\to +\infty}\dfrac{\log(\RWT(w,t))}{t}\right).
\end{equation}
The inner limit exists since $t\mapsto\log\RWT(w,t)$ is sub-additive, and the outer one exists since $\RWT(w,t)$ is non-decreasing in $w$.
The same formula holds in every dimension, with boxes in place of intervals.

\begin{proposition}\label{prop:Ddim}
Let $\glorule$ be a $D$-dimensional CA on a finite alphabet $A$. For $w\geq 0$ let $N_\glorule(w,t)$ be the number of distinct sequences $(\glorule^j(c)|_{B_w})_{0\leq j\leq t}$, $c\in A^\zzd$. Then
\[
\entropy(\glorule)=\lim_{w\to+\infty}\left(\lim_{t\to+\infty}\frac{\log N_\glorule(w,t)}{t}\right).
\]
\end{proposition}
\begin{proof}
Recall that $A^\zzd$ is equipped with the Tychonoff distance $d$ defined in Section~\ref{CAandGCA}; we use the notation of the Bowen--Dinaburg definition with $T=\glorule$.
Fix $w\geq0$ and $t\geq0$. Recall that $d(c,c')<2^{-w}$ if and only if $c$ and $c'$ coincide on $B_w$. Hence, for $x,y\in A^\zzd$,
\[
d_{t+1}(x,y)\geq2^{-w}\iff \glorule^j(x)|_{B_w}\neq\glorule^j(y)|_{B_w}\ \text{for some } 0\leq j\leq t,
\]
that is, $x$ and $y$ are $(t+1,2^{-w})$-separated if and only if the sequences $(\glorule^j(x)|_{B_w})_{0\leq j\leq t}$ and $(\glorule^j(y)|_{B_w})_{0\leq j\leq t}$ are different. Therefore a set $E\subseteq A^\zzd$ is $(t+1,2^{-w})$-separated if and only if distinct points of $E$ give distinct sequences; choosing one configuration for each of the $N_\glorule(w,t)$ possible sequences, we obtain
\begin{equation}\label{eq:sN}
s(t+1,2^{-w})=N_\glorule(w,t).
\end{equation}

We now show that the sequence $a_t=\log N_\glorule(w,t)$ is subadditive. Let $t_1,t_2\geq0$ and $c\in A^\zzd$. The sequence $(\glorule^j(c)|_{B_w})_{0\leq j\leq t_1+t_2}$ is determined by its first part $(\glorule^j(c)|_{B_w})_{0\leq j\leq t_1}$ and by $(\glorule^{i}(c')|_{B_w})_{0\leq i\leq t_2}$, where $c'=\glorule^{t_1}(c)$. Hence $N_\glorule(w,t_1+t_2)\leq N_\glorule(w,t_1)\,N_\glorule(w,t_2)$, \ie, $a_{t_1+t_2}\leq a_{t_1}+a_{t_2}$. Since $a_t\geq0$, by Fekete's lemma the limit $\lim_{t}\frac{a_t}{t}$ exists and is finite. By~\eqref{eq:sN}, and since $\frac{t}{t+1}\to1$,
\begin{equation}\label{eq:limsup}
\limsup_{n\to\infty}\frac1n\log s(n,2^{-w})=\lim_{t\to\infty}\frac{\log s(t+1,2^{-w})}{t+1}=\lim_{t\to\infty}\frac{\log N_\glorule(w,t)}{t}.
\end{equation}

Finally, if $0<\varepsilon'<\varepsilon$, then every $(n,\varepsilon)$-separated set is $(n,\varepsilon')$-separated, so that $s(n,\varepsilon')\geq s(n,\varepsilon)$. Thus $\limsup_n\frac1n\log s(n,\varepsilon)$ is non-increasing in $\varepsilon$, its limit as $\varepsilon\to0$ exists in $[0,+\infty]$, and it can be computed along the sequence $\varepsilon=2^{-w}$, $w\to+\infty$. By~\eqref{eq:limsup} we obtain
\[
\entropy(\glorule)=\lim_{w\to+\infty}\left(\lim_{t\to+\infty}\frac{\log N_\glorule(w,t)}{t}\right).
\]
\end{proof}

For $D=1$, Proposition~\ref{prop:Ddim} gives Equation~\eqref{entropydef} (with windows of width $2w+1$; by monotonicity in the width the two limits coincide).

\section{Topological entropy of one-dimensional GCAs}\label{sec:mainresult}

\noindent Throughout this section $D=1$.

Let $G$ be a finite group with identity element $e$.
Let $\glorule$ be any GCA on $G$ with local rule $f$ and neighborhood $V_f=\{v_1,\dots,v_k\}$. 
Let $l_\glorule$ and $r_\glorule$ be the quantities defined in Equation~\eqref{landr}. 
Let 
\begin{eqnarray}
    L&=&\{ x\in G^\ZZ: \forall i> 0,  x_i=e \} \text{ and}\\
    R&=&\{ x\in G^\ZZ: \forall i< 0,  x_i=e \}.
\end{eqnarray}
Note that $L$ and $R$ are subgroups of $G^\ZZ$. For an interval $I\subseteq\ZZ$ we say that a configuration $c$ is \emph{trivial on $I$} if $c_i=e$ for every $i\in I$.

\begin{figure}[ht]
    \centering
\begin{tikzpicture}[scale=0.8]
\definecolor{lightred}{rgb}{1,0.7,0.7}

 \def\orx{0}
 \def\ory{0}
 \def\width{.4}
 \def\high{.4}
\def\righe{8}
\def\colonne{20}
\pgfmathsetmacro{\altezza}{\high*(\righe+1)}
\pgfmathsetmacro{\larghezza}{\width*(\colonne+1)}

\def\casellay{7}
\pgfmathsetmacro{\cy}{\ory+\casellay*\high}
\def\colore{lightgray}
\foreach \casellax in {0,...,20} {
\pgfmathsetmacro{\cx}{\orx+\casellax*\width}
\filldraw[fill=lightgray, draw=lightgray] (\cx,\cy) rectangle (\cx+\width,\cy+\high);
  }

\foreach \y in {1,...,\righe} {
    \draw (\orx,\ory+\y*\high) -- (\orx+\larghezza,\ory+\y*\high);
  }
\foreach \x in {1,...,\colonne} {
     \draw (\orx+\x*\width,\ory) -- (\orx+\x*\width,\ory+\altezza);
  }

\definecolor{lightlightgray}{gray}{0.95} 
\pgfmathsetmacro{\cy}{\ory+\casellay*\high}
\def\colore{lightgray}
\foreach \casellax in {7,...,13} {
\foreach \casellay in {2,...,7} {
\pgfmathsetmacro{\cx}{\orx+\casellax*\width}
\pgfmathsetmacro{\cy}{\ory+\casellay*\high}
\filldraw[fill=blue!20, draw=none, opacity=0.3] (\cx,\cy) rectangle (\cx+\width,\cy+\high);
  }
}


\def\spessore{very thick}

\draw[black,\spessore] (\orx+7*\width,\ory+2*\high) -- (\orx+14*\width,\ory+2*\high);
\draw[black,\spessore] (\orx+7*\width,\ory+8*\high) -- (\orx+14*\width,\ory+8*\high);
\draw[black,\spessore] (\orx+7*\width,\ory+2*\high) -- (\orx+7*\width,\ory+8*\high);
\draw[black,\spessore] (\orx+14*\width,\ory+2*\high) -- (\orx+14*\width,\ory+8*\high);

 \node at (-0.55,2.56+\high) {\boldsymbol{$x \to$}};
 \node at (3.4-\width*1,+4) {$\downarrow$};
 \node at (3.4-\width*1,4.55) {\boldsymbol{$i$}};
\node at (5.8-\width*1,+4) {$\downarrow$};
 \node at (5.8-\width*1,4.55) {\boldsymbol{$j$}};
 \node at (\orx+\width*24+0.155,2.2) {\boldsymbol{$t+1$}};
 \draw[<->,black,\spessore] (\orx+22.0*\width,\ory+2*\high) -- (\orx+22.0*\width,\ory+8*\high);

\end{tikzpicture}
 \caption{space-time diagram obtained
by iterating $\glorule$ on $x$. The $(t+1)\times (j-i+1)$ rectangle of purple-colored cells represents $\tr_{\glorule,t}(x,i,j)$. 
}
    \label{fig:trace}
\end{figure}
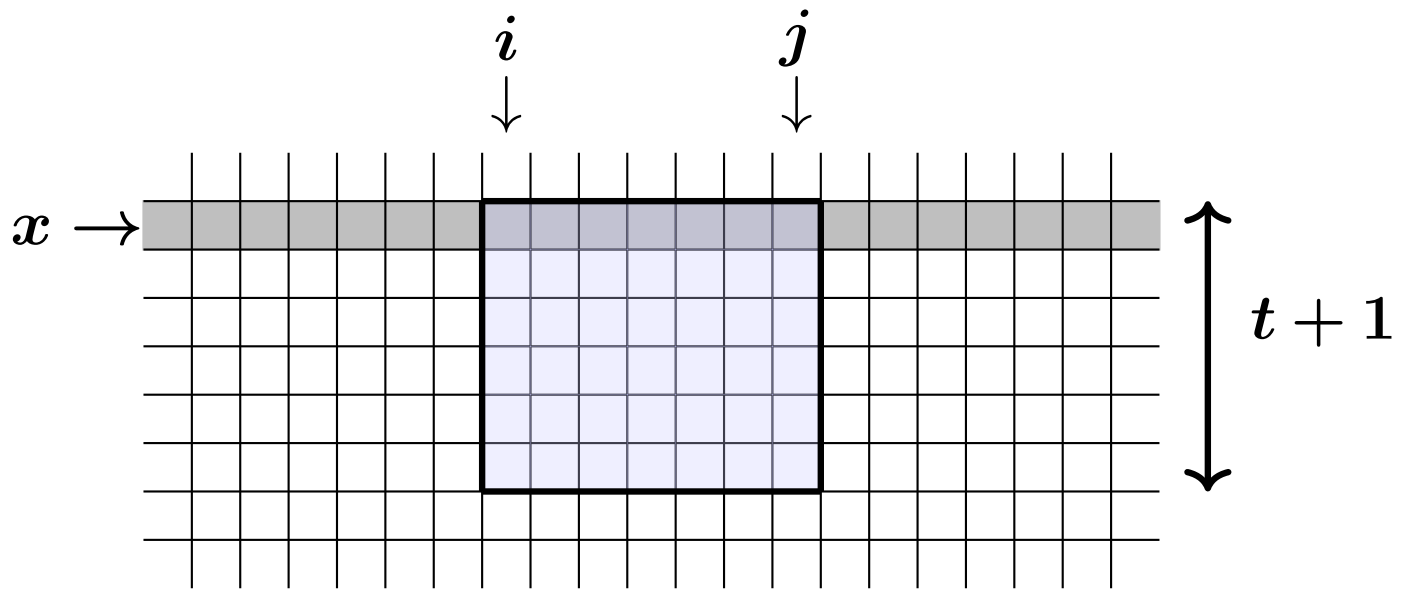

\noindent
For $t\geq0$, $x\in G^\ZZ$ and $i,j\in\ZZ$ with $i\leq j$, let 
$$
\tr_{\glorule,t}(x,i,j)=
\begin{bmatrix}
 x_i & x_{i+1} &\cdots&x_j \\
 \glorule(x)_i& \glorule(x)_{i+1}&\cdots&\glorule(x)_j \\
 \glorule^2(x)_i&\glorule^2(x)_{i+1}&\cdots&\glorule^2(x)_j \\
 \vdots &\vdots &\vdots&\vdots \\
 \glorule^t(x)_i&\glorule^t(x)_{i+1}&\cdots&\glorule^t(x)_j 
 \end{bmatrix}
$$
denote the $(t+1)\times (j-i+1)$ rectangular portion of the space-time diagram obtained
by iterating $\glorule$ on $x$. See Figure~\ref{fig:trace} for a graphical illustration of $\tr_{\glorule,t}(x,i,j)$. 
For $j=i-1$ we let $\tr_{\glorule,t}(x,i,i-1)$ be the empty array; accordingly, every non-empty set of traces of width $0$ has exactly one element. This convention is only needed in the degenerate cases $l_\glorule=0$ or $r_\glorule=0$ below.

\noindent
Since $\glorule$, and hence every iterate $\glorule^s$, is a group endomorphism of $G^\ZZ$, for all $x,y\in G^\ZZ$ we have $\glorule^s(xy)=\glorule^s(x)\glorule^s(y)$ and $\glorule^s(x^{-1})=\glorule^s(x)^{-1}$; therefore
\begin{equation}\label{trhom}
\tr_{\glorule,t}(xy,i,j)=\tr_{\glorule,t}(x,i,j)\cdot\tr_{\glorule,t}(y,i,j)
\quad\text{and}\quad
\tr_{\glorule,t}(x^{-1},i,j)=\tr_{\glorule,t}(x,i,j)^{-1},
\end{equation}
where products and inverses of traces are taken entrywise. In particular, sets of traces of the form $\{\tr_{\glorule,t}(x,i,j): x\in H\}$, with $H$ a subgroup of $G^\ZZ$, are subgroups of $G^{(t+1)\times(j-i+1)}$. Equation~\eqref{trhom} is the point where the group structure enters; it is used in an essential way in
Lemmata~\ref{lem:cancel} and~\ref{lem:welldef}
and in Theorem~\ref{wall1}.

The quantity $\RWT(w,t)$ introduced in the definition of topological entropy (see Equation \eqref{entropydef}) can be expressed as
\begin{equation}\label{R as trace}
    \RWT(w,t)=|\{ 
    \tr_{\glorule,t}(x,k,k+w-1)
    :\ x\in G^\ZZ, k\in \ZZ\}|.
\end{equation}
Since $\glorule$ commutes with $\sigma$, we have $\tr_{\glorule,t}(x,k,k+w-1)=\tr_{\glorule,t}(\sigma^{k-k'}(x),k',k'+w-1)$; hence the set in Equation~\eqref{R as trace} does not change if $k$ is fixed, and $\RWT(w,t)=|\{\tr_{\glorule,t}(x,k,k+w-1): x\in G^\ZZ\}|$ for every fixed $k\in\ZZ$.

\noindent
We now define
\begin{eqnarray}
\lw(\glorule,t)&=&\{
\tr_{\glorule,t}(x,-l_\glorule+1,0)
: x\in L\} 
\text { and}\\
\rw(\glorule,t)&=&\{
\tr_{\glorule,t}(x,0,r_\glorule-1)
: x\in R\}.
 \end{eqnarray}
By Equation~\eqref{trhom}, $\lw(\glorule,t)$ and $\rw(\glorule,t)$ are finite groups. If $l_\glorule=0$ (resp. $r_\glorule=0$), then $|\lw(\glorule,t)|=1$ (resp. $|\rw(\glorule,t)|=1$) by our convention on empty traces.
See Figures~\ref{fig:left} and~\ref{fig:right}  for a graphical illustration of
the elements of $\lw(\glorule,t)$ and $\rw(\glorule,t)$, respectively.

\begin{figure}[ht]
    \centering
\begin{tikzpicture}[scale=0.8]
\definecolor{lightred}{rgb}{1,0.7,0.7}

 \def\orx{0}
 \def\ory{0}
 \def\width{.4}
 \def\high{.4}
\def\righe{8}
\def\colonne{20}
\pgfmathsetmacro{\altezza}{\high*(\righe+1)}
\pgfmathsetmacro{\larghezza}{\width*(\colonne+1)}

\def\casellay{7}
\pgfmathsetmacro{\cy}{\ory+\casellay*\high}
\def\colore{lightgray}
\foreach \casellax in {0,...,10} {
\pgfmathsetmacro{\cx}{\orx+\casellax*\width}
\filldraw[fill=lightgray, draw=lightgray] (\cx,\cy) rectangle (\cx+\width,\cy+\high);
  }

 
\foreach \y in {1,...,\righe} {
    \draw (\orx,\ory+\y*\high) -- (\orx+\larghezza,\ory+\y*\high);
  }
\foreach \x in {1,...,\colonne} {
     \draw (\orx+\x*\width,\ory) -- (\orx+\x*\width,\ory+\altezza);
  }

\definecolor{lightlightgray}{gray}{0.95} 
\pgfmathsetmacro{\cy}{\ory+\casellay*\high}
\def\colore{lightgray}
\foreach \casellax in {6,...,10} {
\foreach \casellay in {2,...,7} {
\pgfmathsetmacro{\cx}{\orx+\casellax*\width}
\pgfmathsetmacro{\cy}{\ory+\casellay*\high}
\filldraw[fill=blue!20, draw=none, opacity=0.3] (\cx,\cy) rectangle (\cx+\width,\cy+\high);
  }
}

\foreach \casellax in {11,...,19} {
  \foreach \casellay in {7,...,7} {
\pgfmathsetmacro{\cx}{\orx+\casellax*\width}
\pgfmathsetmacro{\cy}{\ory+\casellay*\high}
\node at (\cx+.2,\cy+.2) {$e$};

  }
  
}


\def\spessore{very thick}

\draw[black,\spessore] (\orx+6*\width,\ory+2*\high) -- (\orx+11*\width,\ory+2*\high);
\draw[black,\spessore] (\orx+6*\width,\ory+8*\high) -- (\orx+11*\width,\ory+8*\high);
\draw[black,\spessore] (\orx+6*\width,\ory+2*\high) -- (\orx+6*\width,\ory+8*\high);
\draw[black,\spessore] (\orx+11*\width,\ory+2*\high) -- (\orx+11*\width,\ory+8*\high);

 \node at (-0.55,2.56+\high) {\boldsymbol{$x \to$}};
 \node at (4.6-\width*1,+4) {$\downarrow$};
 \node at (4.6-\width*1,4.55) {\boldsymbol{$0$}};
 \node at (\orx+\width*24+0.155,2.2) {\boldsymbol{$t+1$}};
 \draw[<->,black,\spessore] (\orx+22.0*\width,\ory+2*\high) -- (\orx+22.0*\width,\ory+8*\high);

\draw[<->,black,\spessore] (\orx+6.0*\width,\ory+-1*\high) -- (\orx+11.0*\width,\ory+-1*\high);
\node at (\orx+\width*8.5,-1) {\boldsymbol{$l_\glorule$}};

\end{tikzpicture}
 \caption{space-time diagram obtained
by iterating $\glorule$ on $x\in L$. The $(t+1)\times l_\glorule$ rectangle of purple-colored cells represents 
$\tr_{\glorule,t}(x,-l_\glorule+1,0)$ 
which is an element of $\lw(\glorule,t)$
}
    \label{fig:left}
\end{figure}

\begin{figure}[ht]
    \centering
\begin{tikzpicture}[scale=0.8]
\definecolor{lightred}{rgb}{1,0.7,0.7}

  \def\orx{0}
 \def\ory{0}
 \def\width{.4}
 \def\high{.4}
\def\righe{8}
\def\colonne{20}
\pgfmathsetmacro{\altezza}{\high*(\righe+1)}
\pgfmathsetmacro{\larghezza}{\width*(\colonne+1)}

\def\casellay{7}
\pgfmathsetmacro{\cy}{\ory+\casellay*\high}
\def\colore{lightgray}
\foreach \casellax in {10,...,20} {
\pgfmathsetmacro{\cx}{\orx+\casellax*\width}
\filldraw[fill=lightgray, draw=lightgray] (\cx,\cy) rectangle (\cx+\width,\cy+\high);
  }

\foreach \y in {1,...,\righe} {
    \draw (\orx,\ory+\y*\high) -- (\orx+\larghezza,\ory+\y*\high);
  }
\foreach \x in {1,...,\colonne} {
     \draw (\orx+\x*\width,\ory) -- (\orx+\x*\width,\ory+\altezza);
  }

\definecolor{lightlightgray}{gray}{0.95} 
\pgfmathsetmacro{\cy}{\ory+\casellay*\high}
\def\colore{lightgray}
\foreach \casellax in {10,...,12} {
\foreach \casellay in {2,...,7} {
\pgfmathsetmacro{\cx}{\orx+\casellax*\width}
\pgfmathsetmacro{\cy}{\ory+\casellay*\high}
\filldraw[fill=blue!20, draw=none, opacity=0.3] (\cx,\cy) rectangle (\cx+\width,\cy+\high);
  }
}

\foreach \casellax in {1,...,9} {
  \foreach \casellay in {7,...,7} {
\pgfmathsetmacro{\cx}{\orx+\casellax*\width}
\pgfmathsetmacro{\cy}{\ory+\casellay*\high}
\node at (\cx+.2,\cy+.2) {$e$};

  }
  
}


\def\spessore{very thick}

\draw[black,\spessore] (\orx+10*\width,\ory+2*\high) -- (\orx+13*\width,\ory+2*\high);
\draw[black,\spessore] (\orx+10*\width,\ory+8*\high) -- (\orx+13*\width,\ory+8*\high);
\draw[black,\spessore] (\orx+10*\width,\ory+2*\high) -- (\orx+10*\width,\ory+8*\high);
\draw[black,\spessore] (\orx+13*\width,\ory+2*\high) -- (\orx+13*\width,\ory+8*\high);

 \node at (-0.55,2.56+\high) {\boldsymbol{$x \to$}};
 \node at (4.6-\width*1,+4) {$\downarrow$};
 \node at (4.6-\width*1,4.55) {\boldsymbol{$0$}};
 \node at (\orx+\width*24+0.155,2.2) {\boldsymbol{$t+1$}};
 \draw[<->,black,\spessore] (\orx+22.0*\width,\ory+2*\high) -- (\orx+22.0*\width,\ory+8*\high);

\draw[<->,black,\spessore] (\orx+10.0*\width,\ory+-1*\high) -- (\orx+13.0*\width,\ory+-1*\high);
\node at (\orx+\width*11.5,-1) {\boldsymbol{$r_\glorule$}};
\end{tikzpicture}
  \caption{space-time diagram obtained
by iterating $\glorule$ on $x\in R$. The $(t+1)\times r_\glorule$ rectangle of purple-colored cells represents 
$\tr_{\glorule,t}(x,0,r_\glorule-1)$
which is an element of $\rw(\glorule,t)$
}
    \label{fig:right}
\end{figure}

The following two lemmata formalize the fact that information can enter a region of the space-time diagram only through its boundary columns.

\begin{lemma}\label{lem:determinacy}
Let $s\geq 0$.
\begin{itemize}
\item[(i)] Let $U\in L$ and assume that $\glorule^j(U)$ is trivial on $[-l_\glorule+1,0]$ for every $0\leq j<s$. Then $\glorule^j(U)$ is trivial on $[1,+\infty)$ for every $0\leq j\leq s$.
\item[(ii)] Let $\delta\geq-1$ and let $V\in G^\ZZ$ be trivial on $(-\infty,\delta]$. Assume that $\glorule^j(V)$ is trivial on $[\delta+1,\delta+r_\glorule]$ for every $0\leq j<s$. Then $\glorule^j(V)$ is trivial on $(-\infty,\delta]$ for every $0\leq j\leq s$.
\end{itemize}
\end{lemma}
\begin{proof}
(i) We proceed by induction on $j$. For $j=0$ the claim holds since $U\in L$. Let $0\leq j<s$ and assume that $\glorule^j(U)$ is trivial on $[1,+\infty)$. For $i\geq1$ we have $\glorule^{j+1}(U)_i=f(\glorule^j(U)_{i+v_1},\dots,\glorule^j(U)_{i+v_k})$. Since $-l_\glorule\leq v_h\leq r_\glorule$, every index $i+v_h$ satisfies $i+v_h\geq 1-l_\glorule$, hence it belongs either to $[-l_\glorule+1,0]$, where $\glorule^j(U)$ is trivial by hypothesis (because $j<s$), or to $[1,+\infty)$, where $\glorule^j(U)$ is trivial by the inductive hypothesis. Since $f$ is a homomorphism, $f(e,\dots,e)=e$, and thus $\glorule^{j+1}(U)_i=e$.

(ii) The proof is symmetric: for $i\leq\delta$ every index $i+v_h$ satisfies $i+v_h\leq\delta+r_\glorule$, hence it belongs either to $[\delta+1,\delta+r_\glorule]$ or to $(-\infty,\delta]$.
\end{proof}

\begin{lemma}\label{lem:cancel}
Let $\delta\geq0$ and $t\geq0$. Let $U\in L$ and let $V\in G^\ZZ$ be trivial on $(-\infty,\delta]$. If
\[
\tr_{\glorule,t}(U,-l_\glorule+1,r_\glorule+\delta)=\tr_{\glorule,t}(V,-l_\glorule+1,r_\glorule+\delta),
\]
then, for every $0\leq j\leq t$, $\glorule^j(U)$ is trivial on $[-l_\glorule+1,0]$ and $\glorule^j(V)$ is trivial on $[\delta+1,\delta+r_\glorule]$.
\end{lemma}
\begin{proof}
We proceed by induction on $j\in\{0,\dots,t\}$. Assume that the claim holds for every $j'<j$ (no assumption is needed for $j=0$). By Lemma~\ref{lem:determinacy}(i) with $s=j$, $\glorule^j(U)$ is trivial on $[1,+\infty)$, which contains $[\delta+1,\delta+r_\glorule]$ since $\delta\geq0$. By Lemma~\ref{lem:determinacy}(ii) with $s=j$, $\glorule^j(V)$ is trivial on $(-\infty,\delta]$, which contains $[-l_\glorule+1,0]$. Since $\glorule^j(U)$ and $\glorule^j(V)$ coincide on $[-l_\glorule+1,r_\glorule+\delta]\supseteq[-l_\glorule+1,0]\cup[\delta+1,\delta+r_\glorule]$, the claim holds for $j$.
\end{proof}

\begin{figure}[ht]
    \centering
\begin{tikzpicture}[scale=0.8]
\definecolor{lightred}{rgb}{1,0.7,0.7}

 \def\orx{0}
 \def\ory{0}
 \def\width{.4}
 \def\high{.4}
\def\righe{20}
\def\colonne{30}
\pgfmathsetmacro{\altezza}{\high*(\righe+1)}
\pgfmathsetmacro{\larghezza}{\width*(\colonne+1)}

\def\casellay{7}
\pgfmathsetmacro{\cy}{\ory+\casellay*\high}
\def\colore{lightgray}

\foreach \y in {1,...,\righe} {
    \draw (\orx,\ory+\y*\high) -- (\orx+\larghezza,\ory+\y*\high);
  }
\foreach \x in {1,...,\colonne} {
     \draw (\orx+\x*\width,\ory) -- (\orx+\x*\width,\ory+\altezza);
  }

\definecolor{lightlightgray}{gray}{0.95} 
\pgfmathsetmacro{\cy}{\ory+\casellay*\high}
\def\colore{lightgray}

\foreach \casellax in {3,...,4} {
\foreach \casellay in {2,...,19} {
\pgfmathsetmacro{\cx}{\orx+\casellax*\width}
\pgfmathsetmacro{\cy}{\ory+\casellay*\high}
\filldraw[fill=yellow!100, draw=none, opacity=0.3] (\cx,\cy) rectangle (\cx+\width,\cy+\high);
  }
}

\foreach \casellax in {25,...,27} {
\foreach \casellay in {2,...,19} {
\pgfmathsetmacro{\cx}{\orx+\casellax*\width}
\pgfmathsetmacro{\cy}{\ory+\casellay*\high}
\filldraw[fill=green!100, draw=none, opacity=0.3] (\cx,\cy) rectangle (\cx+\width,\cy+\high);
  }
}

\foreach \casellax in {5,...,24} {
\foreach \casellay in {2,...,19} {
\pgfmathsetmacro{\cx}{\orx+\casellax*\width}
\pgfmathsetmacro{\cy}{\ory+\casellay*\high}
\filldraw[fill=red!80, draw=none, opacity=0.3] (\cx,\cy) rectangle (\cx+\width,\cy+\high);
  }
}

\foreach \casellax in {5,...,24} {
  \foreach \casellay in {19,...,19} {
\pgfmathsetmacro{\cx}{\orx+\casellax*\width}
\pgfmathsetmacro{\cy}{\ory+\casellay*\high}
\node at (\cx+.2,\cy+.2) {$e$};

  }
  
}

\def\spessore{thick}
\draw[->,black,\spessore,dashed] (\orx+4.5*\width,\ory+19.5*\high) -- (\orx+30.5*\width,\ory+6.5*\high);

\draw[->,black,\spessore,dashed] (\orx+25.5*\width,\ory+19.5*\high) -- (\orx+1.5*\width,\ory+11.5*\high);

\def\spessore{very thick}
\draw[black,\spessore] (\orx+3*\width,\ory+20*\high) -- (\orx+28*\width,\ory+20*\high);
\draw[black,\spessore] (\orx+3*\width,\ory+2*\high) -- (\orx+28*\width,\ory+2*\high);
\draw[black,\spessore] (\orx+3*\width,\ory+2*\high) -- (\orx+3*\width,\ory+20*\high);
\draw[black,\spessore] (\orx+28*\width,\ory+2*\high) -- (\orx+28*\width,\ory+20*\high);

 \node at (\orx+\width*34+0.155,4.2) {\boldsymbol{$t+1$}};
 \draw[<->,black,\spessore] (\orx+32.0*\width,\ory+2*\high) -- (\orx+32.0*\width,\ory+20*\high);

\draw[<->,black,\spessore] (\orx+3.0*\width,\ory+-1*\high) -- (\orx+5.0*\width,\ory+-1*\high);
\node at (\orx+\width*4,-1) {\boldsymbol{$l_\glorule$}};

\draw[<->,black,\spessore] (\orx+5*\width,\ory+-1*\high) -- (\orx+25*\width,\ory+-1*\high);
\node at (\orx+\width*15,-1) {\boldsymbol{$\delta$}};

\draw[<->,black,\spessore] (\orx+25.0*\width,\ory+-1*\high) -- (\orx+28.0*\width,\ory+-1*\high);
\node at (\orx+\width*26.5,-1) {\boldsymbol{$r_\glorule$}};

\end{tikzpicture}
  \caption{
Graphical representation of $a \opdelta b$, obtained from an element
$a \in \lw(\glorule,t)$ (yellow-colored cells on the left) and an element
$b \in \rw(\glorule,t)$ (green-colored cells on the right).
The trace $a \opdelta b$ is the $(t+1)\times (l_\glorule + \delta + r_\glorule)$
rectangular box delimited by the thick black border.
Yellow-colored cells below the dashed arrow can be influenced by green-colored cells on the right and, symmetrically, green-colored cells below the dashed arrow can be influenced by yellow-colored cells on the left.
}
 \label{fig:leftright}
\end{figure}
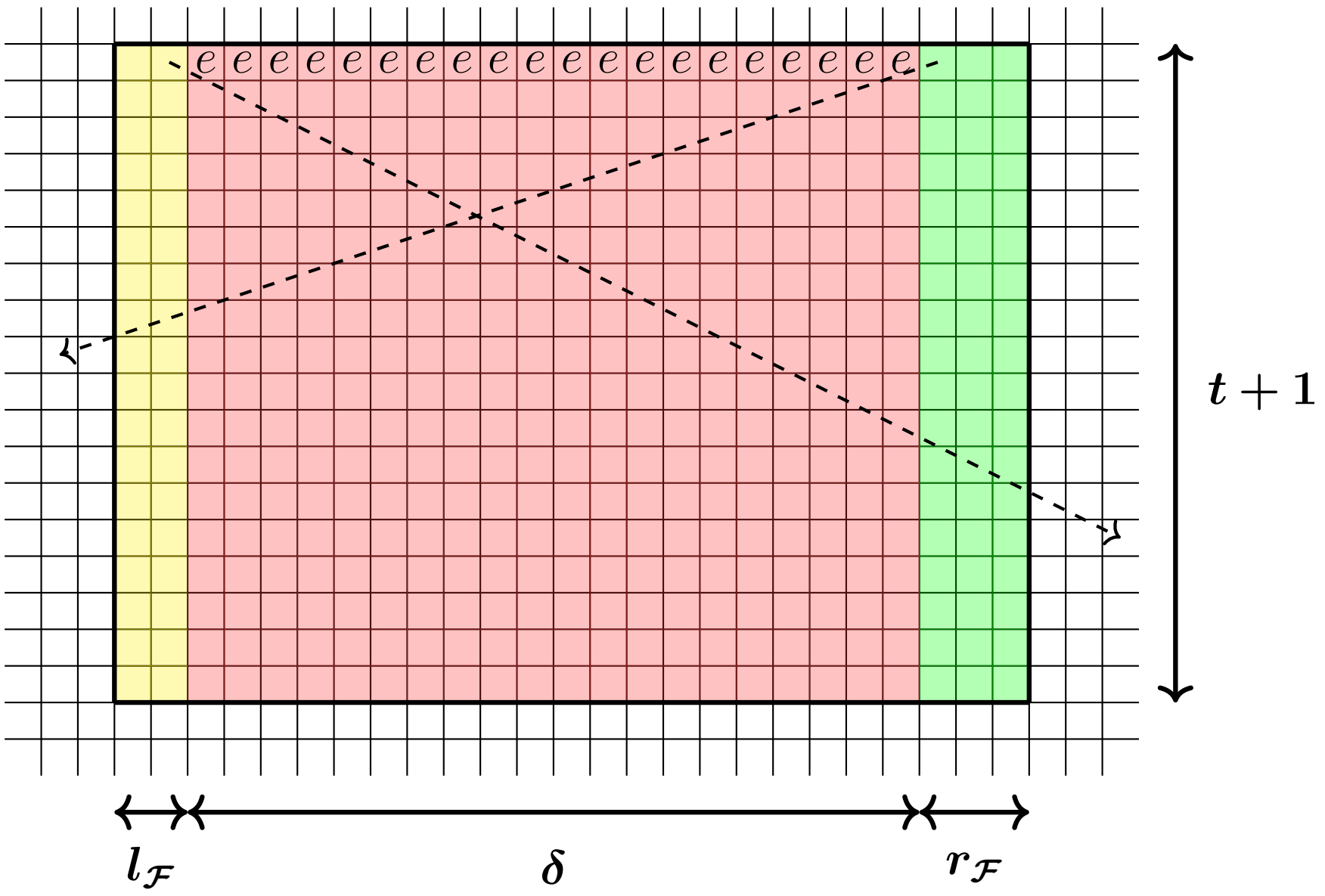

Let $\delta$ be any non-negative integer, let $a\in \lw(\glorule,t)$ and $b\in \rw(\glorule,t)$. There is a natural way to define a new trace, denoted by $a\opdelta b$, as follows (see Figure~\ref{fig:leftright}).
Let $X\in L$ be such that $\tr_{\glorule,t}(X,-l_\glorule+1,0)=a$ and $Y\in R$ be such that $\tr_{\glorule,t}(Y,0,r_\glorule-1)=b$. Then
$$a\opdelta b=\tr_{\glorule,t}\big(X\,\rho^{\delta+1}(Y),-l_\glorule+1,r_\glorule+\delta\big).$$
Note that $X$ is supported in $(-\infty,0]$, $\rho^{\delta+1}(Y)$ is supported in $[\delta+1,+\infty)$, and $\rho^{\delta+1}(Y)$ has the trace $b$ on the columns $[\delta+1,\delta+r_\glorule]$.

\begin{lemma}\label{lem:welldef}
The trace $a\opdelta b$ does not depend on the choice of $X$ and $Y$.
\end{lemma}
\begin{proof}
Let $X,X'\in L$ with $\tr_{\glorule,t}(X,-l_\glorule+1,0)=\tr_{\glorule,t}(X',-l_\glorule+1,0)=a$. Then $U=X'^{-1}X\in L$ and, by Equation~\eqref{trhom}, $\glorule^j(U)$ is trivial on $[-l_\glorule+1,0]$ for every $j\leq t$. By Lemma~\ref{lem:determinacy}(i) with $s=t+1$, $\glorule^j(U)$ is trivial on $[1,+\infty)$ for every $j\leq t$. Hence $\glorule^j(X)=\glorule^j(X')$ on $[-l_\glorule+1,+\infty)$ for every $j\leq t$. In the same way, if $Y,Y'\in R$ have the same trace $b$, then $V=\rho^{\delta+1}(Y'^{-1}Y)$ is trivial on $(-\infty,\delta]$ and, since $\glorule$ commutes with $\rho$, $\glorule^j(V)$ is trivial on $[\delta+1,\delta+r_\glorule]$ for every $j\leq t$; by Lemma~\ref{lem:determinacy}(ii) with $s=t+1$, $\glorule^j(V)$ is trivial on $(-\infty,\delta]$ for every $j\leq t$, so that $\glorule^j(\rho^{\delta+1}(Y))=\glorule^j(\rho^{\delta+1}(Y'))$ on $(-\infty,\delta+r_\glorule]$ for every $j\leq t$. Since $[-l_\glorule+1,r_\glorule+\delta]$ is contained in both $[-l_\glorule+1,+\infty)$ and $(-\infty,\delta+r_\glorule]$, Equation~\eqref{trhom} gives $\tr_{\glorule,t}(X\rho^{\delta+1}(Y),-l_\glorule+1,r_\glorule+\delta)=\tr_{\glorule,t}(X'\rho^{\delta+1}(Y'),-l_\glorule+1,r_\glorule+\delta)$.
\end{proof}

Graphically, as it can be seen in Figure~\ref{fig:leftright}, the red-colored region of the space-time diagram depends solely on the yellow-colored region (on the left) and on the green-colored region (on the right).
Note that the first $l_\glorule$ columns of $a\opdelta b$ and the last $r_\glorule$ columns of $a\opdelta b$ will in general differ from $a$ and $b$,
 respectively. This is due to the fact that 
$a$ and $b$ may influence each other when 
$t$ is sufficiently large.

\noindent
We define
\begin{eqnarray}
\w(\glorule,t,\delta)&=&\{a\opdelta b:\ a\in \lw(\glorule,t),\ b\in \rw(\glorule,t)  \}. \label{walldef}
\end{eqnarray}

\begin{theorem}\label{wall1} 
For every $\delta \geq0$ and $t\geq 0$ we have
$$|\lw(\glorule,t)|\, |\rw(\glorule,t)|= |\w(\glorule,t,\delta)|.$$
\end{theorem}
\begin{proof} 
By Equation~\eqref{walldef} we have $|\w(\glorule,t,\delta)|\leq |\lw(\glorule,t)|\, |\rw(\glorule,t)|$, so it suffices to prove that the map $(a,b)\mapsto a\opdelta b$ is injective. Note that, a priori, the space--time diagrams generated by $a$ and $b$ might interfere with each other after a sufficiently large number of iterations, so that different pairs could produce the same trace (see Figure~\ref{fig:leftright}). We show that the group structure rules this out.

Let $a_1,a_2\in\lw(\glorule,t)$ and $b_1,b_2\in\rw(\glorule,t)$ with $a_1\opdelta b_1=a_2\opdelta b_2$. Choose $X_1,X_2\in L$ and $Y_1,Y_2\in R$ with $\tr_{\glorule,t}(X_h,-l_\glorule+1,0)=a_h$ and $\tr_{\glorule,t}(Y_h,0,r_\glorule-1)=b_h$ for $h=1,2$, and write $W=[-l_\glorule+1,r_\glorule+\delta]$. For every $j\leq t$ and $i\in W$ we have
\[
\glorule^j(X_1)_i\,\glorule^j(\rho^{\delta+1}(Y_1))_i=\glorule^j(X_2)_i\,\glorule^j(\rho^{\delta+1}(Y_2))_i ,
\]
and hence
\[
\glorule^j(X_2)_i^{-1}\,\glorule^j(X_1)_i=\glorule^j(\rho^{\delta+1}(Y_2))_i\,\glorule^j(\rho^{\delta+1}(Y_1))_i^{-1}.
\]
Since $\glorule^j$ is an endomorphism commuting with $\rho$, this means that $U=X_2^{-1}X_1\in L$ and $V=\rho^{\delta+1}(Y_2Y_1^{-1})$, which is trivial on $(-\infty,\delta]$, have the same trace on $W$. By Lemma~\ref{lem:cancel}, $\glorule^j(U)$ is trivial on $[-l_\glorule+1,0]$ and $\glorule^j(V)$ is trivial on $[\delta+1,\delta+r_\glorule]$ for every $j\leq t$. By Equation~\eqref{trhom}, the first fact gives $a_1=\tr_{\glorule,t}(X_2U,-l_\glorule+1,0)=a_2$, and the second one gives $\tr_{\glorule,t}(Y_2Y_1^{-1},0,r_\glorule-1)=e$, that is, $b_1=b_2$.
 \end{proof}

The groups $\lw(\glorule,t)$ and $\rw(\glorule,t)$ have a simple structure as $t$ varies. For $t\geq0$ let
\begin{eqnarray*}
N^-_t&=&\{\glorule^{t+1}(U)_{[-l_\glorule+1,0]}:\ U\in L,\ \glorule^j(U) \text{ is trivial on } [-l_\glorule+1,0] \text{ for } 0\leq j\leq t\},\\
N^+_t&=&\{\glorule^{t+1}(U)_{[0,r_\glorule-1]}:\ U\in R,\ \glorule^j(U) \text{ is trivial on } [0,r_\glorule-1] \text{ for } 0\leq j\leq t\}.
\end{eqnarray*}
They are subgroups of $G^{l_\glorule}$ and of $G^{r_\glorule}$, respectively, being images of subgroups of $G^\ZZ$ under homomorphisms.

\begin{proposition}\label{prop:structure}
For every $t\geq0$:
\begin{itemize}
\item[(i)] $|\lw(\glorule,t+1)|=|\lw(\glorule,t)|\,|N^-_t|$ and $|\rw(\glorule,t+1)|=|\rw(\glorule,t)|\,|N^+_t|$; moreover $|\lw(\glorule,0)|=|G|^{l_\glorule}$ and $|\rw(\glorule,0)|=|G|^{r_\glorule}$;
\item[(ii)] $N^-_{t+1}\leq N^-_t$ and $N^+_{t+1}\leq N^+_t$.
\end{itemize}
Consequently there exist $t_0\geq 0$ and subgroups $N^-_\infty\leq G^{l_\glorule}$, $N^+_\infty\leq G^{r_\glorule}$ such that $N^\pm_t=N^\pm_\infty$ for every $t\geq t_0$, and
\[
\lim_{t\to\infty}\frac{\log|\lw(\glorule,t)|}{t}=\log|N^-_\infty|,
\qquad
\lim_{t\to\infty}\frac{\log|\rw(\glorule,t)|}{t}=\log|N^+_\infty|.
\]
\end{proposition}
\begin{proof}
We prove the statements for $\lw$; those for $\rw$ are symmetric, using Lemma~\ref{lem:determinacy}(ii) with $\delta=-1$.

(i) The restriction to the first $t+1$ rows is a surjective group homomorphism $\pi_t:\lw(\glorule,t+1)\to\lw(\glorule,t)$. Its kernel consists of the traces $\tr_{\glorule,t+1}(U,-l_\glorule+1,0)$ with $U\in L$ and $\glorule^j(U)$ trivial on $[-l_\glorule+1,0]$ for $j\leq t$; such a trace is determined by its last row, so the map sending it to its last row is an isomorphism between $\Ker(\pi_t)$ and $N^-_t$. Hence $|\lw(\glorule,t+1)|=|\lw(\glorule,t)|\,|N^-_t|$. Finally $|\lw(\glorule,0)|=|G|^{l_\glorule}$, since the values of $U\in L$ on $[-l_\glorule+1,0]$ are arbitrary.

(ii) Let $g\in N^-_{t+1}$, say $g=\glorule^{t+2}(U)_{[-l_\glorule+1,0]}$ with $U\in L$ and $\glorule^j(U)$ trivial on $[-l_\glorule+1,0]$ for $j\leq t+1$. By Lemma~\ref{lem:determinacy}(i) with $s=t+2$, $\glorule(U)$ is trivial on $[1,+\infty)$, \ie, $U'=\glorule(U)\in L$. Moreover $\glorule^j(U')=\glorule^{j+1}(U)$ is trivial on $[-l_\glorule+1,0]$ for $j\leq t$, and $\glorule^{t+1}(U')_{[-l_\glorule+1,0]}=g$. Hence $g\in N^-_t$.

A decreasing chain of subgroups of the finite group $G^{l_\glorule}$ is eventually constant, which gives $t_0$ and $N^-_\infty$. By (i), $\log|\lw(\glorule,t)|=l_\glorule\log|G|+\sum_{j=0}^{t-1}\log|N^-_j|$, and dividing by $t$ we get the limit.
\end{proof}

By Proposition~\ref{prop:structure} the following quantities are well defined:
\begin{equation}
\entropy^- (\glorule)=
\lim_{t \to \infty} \frac{\log(|\lw(\glorule,t)|)}{t} 
\text{ \  and  \ }
\entropy^+ (\glorule)=\lim_{t \to \infty} \frac{\log(|\rw(\glorule,t)|)}{t}. 
\end{equation}

\noindent
Let $\delta$ be any non-negative integer and
$$C(\delta)=\{ x \in G^\ZZ: x_i=e \text{ for every } i\leq0 \text{ and every } i>\delta \}$$
(in particular $C(0)=\{e^\ZZ\}$). We define
\begin{equation}\label{tstar}
\w^*(\glorule,t,\delta)= \{\tr_{\glorule,t}\big(X\,h\,\rho^{\delta+1}(Y),-l_\glorule+1,r_\glorule+\delta\big):\ X\in L,\ Y\in R,\ h\in C(\delta)  \}.
\end{equation}
Looking at Figure~\ref{fig:leftright}, the configuration $h$ allows one to replace the sequence of $e$'s (first row colored in red) with a sequence of $\delta$ arbitrary elements of $G$.

\begin{lemma}\label{lem:tstar}
For every $\delta\geq0$ and $t\geq0$:
\begin{itemize}
\item[(i)] $|\w^*(\glorule,t,\delta)|=\RWT(l_\glorule+\delta+r_\glorule,t)$;
\item[(ii)] $|\w(\glorule,t,\delta)|\leq|\w^*(\glorule,t,\delta)|\leq|G|^\delta\,|\w(\glorule,t,\delta)|$.
\end{itemize}
\end{lemma}
\begin{proof}
(i) By Equation~\eqref{R as trace} and the subsequent remark with $k=-l_\glorule+1$, $\RWT(l_\glorule+\delta+r_\glorule,t)$ is the number of traces $\tr_{\glorule,t}(c,-l_\glorule+1,r_\glorule+\delta)$, $c\in G^\ZZ$. Every element of $\w^*(\glorule,t,\delta)$ is such a trace. Conversely, given $c\in G^\ZZ$, let $X\in L$ coincide with $c$ on $(-\infty,0]$, let $h\in C(\delta)$ coincide with $c$ on $[1,\delta]$, and let $Y=\sigma^{\delta+1}(c')$, where $c'$ coincides with $c$ on $[\delta+1,+\infty)$ and is trivial elsewhere. Then $Y\in R$ and, since $X$, $h$ and $\rho^{\delta+1}(Y)$ have pairwise disjoint supports, $c=X\,h\,\rho^{\delta+1}(Y)$.

(ii) Taking $h=e^\ZZ$ shows that $\w(\glorule,t,\delta)\subseteq\w^*(\glorule,t,\delta)$. Conversely, since $X$, $h$ and $\rho^{\delta+1}(Y)$ have pairwise disjoint supports, they commute, and by Equation~\eqref{trhom}
$$
\tr_{\glorule,t}\big(X\,h\,\rho^{\delta+1}(Y),-l_\glorule+1,r_\glorule+\delta\big)=$$
$$\tr_{\glorule,t}\big(X\,\rho^{\delta+1}(Y),-l_\glorule+1,r_\glorule+\delta\big)\cdot\tr_{\glorule,t}(h,-l_\glorule+1,r_\glorule+\delta).
$$
The first factor belongs to $\w(\glorule,t,\delta)$, and the second one is determined by $h_{[1,\delta]}\in G^\delta$.
\end{proof}

\begin{theorem}\label{hat} Let $\glorule$ be a GCA on a finite group $G$. Then, for every $\delta\geq0$,
\begin{equation}\label{qualunquedelta}
\entropy(\glorule)=\lim_{t \to \infty} \frac{\log(|\w(\glorule,t,\delta)|)}{t}=\lim_{t \to \infty} \frac{\log\RWT(l_\glorule+\delta+r_\glorule,t)}{t},
\end{equation}
and
\begin{equation}\label{thm1}
\entropy (\glorule)= \entropy^- (\glorule)+\entropy^+ (\glorule).
\end{equation}
\end{theorem}
\begin{proof} 
Fix $\delta\geq0$. By Theorem~\ref{wall1} and Proposition~\ref{prop:structure},
\[
\lim_{t\to\infty}\frac{\log|\w(\glorule,t,\delta)|}{t}=\lim_{t\to\infty}\frac{\log|\lw(\glorule,t)|+\log|\rw(\glorule,t)|}{t}=\entropy^-(\glorule)+\entropy^+(\glorule),
\]
and this value does not depend on $\delta$. By Lemma~\ref{lem:tstar},
\[
\frac{\log|\w(\glorule,t,\delta)|}{t}\leq\frac{\log\RWT(l_\glorule+\delta+r_\glorule,t)}{t}\leq\frac{\log|\w(\glorule,t,\delta)|}{t}+\frac{\delta\log|G|}{t},
\]
so that $\lim_{t}\frac1t\log\RWT(l_\glorule+\delta+r_\glorule,t)=\entropy^-(\glorule)+\entropy^+(\glorule)$ for every $\delta\geq0$. Since the outer limit in Equation~\eqref{entropydef} exists, it can be computed along the subsequence $w=l_\glorule+r_\glorule+\delta$, $\delta\to\infty$, and we obtain $\entropy(\glorule)=\entropy^-(\glorule)+\entropy^+(\glorule)$. This proves both Equation~\eqref{thm1} and Equation~\eqref{qualunquedelta}.
\end{proof}

\medskip\noindent
An alternative interpretation of Theorem~\ref{hat}, obtained by taking $\delta=0$ in Equation~\eqref{qualunquedelta}, is given by the following equation:
\begin{equation}\label{entropydef2}
    \entropy (\glorule)=\lim_{w\to +\infty}\left( \lim_{t\to +\infty}\dfrac{\log(\RWT(w,t))}{t}\right)= \lim_{t\to +\infty}\dfrac{\log(\RWT(l_\glorule+r_\glorule,t))}{t}.
\end{equation}
In other words, for GCAs the topological entropy is already detected by space-time windows of the fixed width $l_\glorule+r_\glorule$.

\begin{corollary}\label{cor:logint}
Let $\glorule$ be a GCA on a finite group $G$. Then $\entropy(\glorule)=\log\big(|N^-_\infty|\,|N^+_\infty|\big)$. In particular, $\entropy(\glorule)=\log k$ for some positive integer $k$ dividing $|G|^{l_\glorule+r_\glorule}$, and $\entropy(\glorule)\leq(l_\glorule+r_\glorule)\log|G|$.
\end{corollary}
\begin{proof}
The first statement follows from Theorem~\ref{hat} and Proposition~\ref{prop:structure}. Since $N^-_\infty\leq G^{l_\glorule}$ and $N^+_\infty\leq G^{r_\glorule}$, Lagrange's theorem gives the rest.
\end{proof}

\begin{theorem}\label{computable} 
There is an algorithm which, given a finite group $G$ by its multiplication table and the local rule of a GCA $\glorule$ on $G$, computes $\entropy(\glorule)$.
\end{theorem}

\begin{proof}
Let $w=l_\glorule+r_\glorule$. If $w=0$, then $\entropy(\glorule)=0$ by Corollary~\ref{cor:logint}, so we may assume $w\geq1$, and we let $I=\{0,\dots,w-1\}$. We use the terminology of~\cite{BeaurK24}. A bi-infinite orbit of $\glorule$ is a sequence $(c^{(s)})_{s\in\ZZ}$ of configurations with $c^{(s+1)}=\glorule(c^{(s)})$ for every $s\in\ZZ$; the \emph{limit set} $\Omega_\glorule$ is the set of configurations lying on some bi-infinite orbit, and the \emph{$D$-trace subshift} $\mathrm{Tr}_I(\glorule)\subseteq(G^I)^\ZZ$ is the set of sequences $(c^{(s)}|_I)_{s\in\ZZ}$, where $(c^{(s)})_{s\in\ZZ}$ ranges over the bi-infinite orbits of $\glorule$. By~\cite[Lemma~14]{BeaurK24}, $\mathrm{Tr}_I(\glorule)$ is a one-dimensional group shift over $G^I$, and by~\cite[Lemma~15]{BeaurK24} it can be effectively constructed from the local rule of $\glorule$, \ie, one can compute a finite set of forbidden words defining it as a subshift of finite type (recall that group shifts are of finite type~\cite{Kitchens1987,KitchensSchmidt1989}, see also~\cite[Theorem~1]{BeaurK24}). Let $\mathcal B_m$ denote the set of words of length $m$ occurring in $\mathrm{Tr}_I(\glorule)$.

We claim that $\entropy(\glorule)=\lim_{m\to\infty}\frac1m\log|\mathcal B_m|$, the entropy of the subshift $\mathrm{Tr}_I(\glorule)$. By~\cite[Lemma~18]{BeaurK24} there is $k\geq0$ such that $\glorule^k(G^\ZZ)=\Omega_\glorule$. Recall that $\RWT(w,t)$ is the number of traces $\tr_{\glorule,t}(c,0,w-1)$, $c\in G^\ZZ$, \ie, of the words $(\glorule^s(c)|_I)_{0\leq s\leq t}$. Every word in $\mathcal B_{t+1}$ is of this form, with $c$ an element of a bi-infinite orbit, so that $|\mathcal B_{t+1}|\leq\RWT(w,t)$. Conversely, for $t\geq k$, the word $(\glorule^s(c)|_I)_{0\leq s\leq t}$ is determined by its first $k$ letters, for which there are at most $|G|^{wk}$ possibilities, and by the word $(\glorule^{s}(c')|_I)_{0\leq s\leq t-k}$ with $c'=\glorule^k(c)\in\Omega_\glorule$; since $c'$ lies on a bi-infinite orbit, the latter word belongs to $\mathcal B_{t+1-k}$. Hence
\[
|\mathcal B_{t+1}|\ \leq\ \RWT(w,t)\ \leq\ |G|^{wk}\,|\mathcal B_{t+1-k}|\qquad(t\geq k).
\]
Dividing the logarithms by $t$ and letting $t\to\infty$, Equation~\eqref{entropydef2} gives the claim.

Finally, the entropy of a one-dimensional subshift of finite type given by a finite set of forbidden words is the logarithm of the spectral radius of an effectively computable non-negative integer matrix (the adjacency matrix of a suitable higher block presentation, see~\cite[Chapter~4]{LindMarcus1995}), and it is therefore computable.
\end{proof}

A second, independent, proof of Theorem~\ref{computable}, which also yields an explicit formula, is given in Remark~\ref{rem:secondproof} below.






\section{Linear GCAs over $(\ZZ/p\ZZ)^n$: an explicit formula via Pontryagin duality}\label{sec:linear}

In this section the dimension $D\geq1$ is arbitrary. Throughout the section $p$ is a prime, $\Fp=\ZZ/p\ZZ$, $n\geq1$, and $\LP=\Fp[x_1^{\pm1},\dots,x_D^{\pm1}]$ denotes the ring of Laurent polynomials in $D$ variables with coefficients in $\Fp$. For $u\in\zzd$ we write $x^u=x_1^{u_1}\cdots x_D^{u_D}$, and for $0\neq q=\sum_u q_ux^u\in\LP$ we let $\supp(q)=\{u\in\zzd:\ q_u\neq0\}$. In this section the letter $t$ denotes an indeterminate.

\medskip\noindent
\textbf{Linear GCAs and their matrices.}
Every endomorphism of the elementary abelian group $\Fp^n$ is $\Fp$-linear. Hence the local rule of a $D$-GCA $\glorule$ on $\Fp^n$ has the form $f(g_1,\dots,g_k)=\sum_{h=1}^k A_{v_h}g_h$ with $A_{v_h}\in M_n(\Fp)$, \ie,
\[
\glorule(c)_v=\sum_{u\in V_f}A_u\,c_{v+u}\qquad (c\in(\Fp^n)^\zzd,\ v\in\zzd).
\]
We associate with $\glorule$ the matrix
\[
M_\glorule(x)=\sum_{u\in V_f}A_u\,x^{-u}\in M_n(\LP).
\]
Identifying a finite configuration $c$ with the vector of Laurent polynomials $\sum_v c_v x^v\in\LP^n$, the map $\glorule$ acts on finite configurations as the multiplication by $M_\glorule$; since finite configurations are dense and CAs are continuous, it follows that $M_{\glorule\circ\glorule'}=M_\glorule M_{\glorule'}$ and $M_{\glorule^k}=M_\glorule^k$. Other conventions used in the literature replace $x_i$ by $x_i^{-1}$; all the statements of this section are invariant under this change.
For $M\in M_n(\LP)$ we write
\[
\chi_M(t)=\det(tI-M)=t^n+c_{n-1}t^{n-1}+\cdots+c_1t+c_0,\qquad c_i\in\LP,
\]
for its characteristic polynomial. When $D=1$ we set
\begin{align*}
E^+_M&=\max\big(\{0\}\cup\{\max\supp(c_i):\ c_i\neq0\}\big),\\
E^-_M&=\min\big(\{0\}\cup\{\min\supp(c_i):\ c_i\neq0\}\big).
\end{align*}
In other words, $[E^-_M,E^+_M]$ is the smallest interval containing $0$ and all the exponents of $x$ appearing in the coefficients of $\chi_M$.

\begin{theorem}\label{formula}
Let $\glorule$ be a $D$-GCA on $\Fp^n$ and let $M=M_\glorule$, with characteristic polynomial $\chi_M(t)=t^n+c_{n-1}t^{n-1}+\cdots+c_0$.
\begin{itemize}
\item[(1)] If $D=1$, then $\entropy(\glorule)=(E^+_M-E^-_M)\log p$.
\item[(2)] If $D\geq2$, then $\entropy(\glorule)=0$ if $c_0,\dots,c_{n-1}\in\Fp$, and $\entropy(\glorule)=+\infty$ otherwise.
\item[(3)] In every dimension, $\glorule$ is equicontinuous if and only if $c_0,\dots,c_{n-1}\in\Fp$.
\end{itemize}
\end{theorem}

An alternative proof of parts (2) and (3), which avoids Pontryagin duality and only uses the one-dimensional results of Section~\ref{sec:mainresult}, is given in~\ref{app:noduality}.
The proof below relies on Pontryagin duality, on the Bridge Theorem relating topological and algebraic entropy, and on a computation of algebraic entropy in terms of ranks of modules over $\Fp[t]$, following the approach of~\cite{ADGT2024}, where finitary linear cellular automata over $\ZZ/m\ZZ$ with $n=1$ and $D=1$ are studied.

\medskip\noindent
\textbf{Algebraic entropy and the Bridge Theorem.}
Let $A$ be an abelian group and $\varphi$ an endomorphism of $A$. For a finite subgroup $E$ of $A$ and $m\geq1$ let $T_m(\varphi,E)=E+\varphi(E)+\cdots+\varphi^{m-1}(E)$. The \emph{algebraic entropy} of $\varphi$ is
\[
\entropy_{alg}(\varphi)=\sup_E\ \lim_{m\to\infty}\frac{\log|T_m(\varphi,E)|}{m},
\]
where $E$ ranges over the finite subgroups of $A$ (the limit exists, see~\cite{DikranjanGiordanoBruno2012}). For a compact abelian group $X$ we denote by $\widehat X$ its Pontryagin dual, \ie, the discrete group of continuous homomorphisms $X\to\mathbb T=\mathbb R/\ZZ$, and for a continuous endomorphism $\psi$ of $X$ we denote by $\widehat\psi:\widehat X\to\widehat X$, $\chi\mapsto\chi\circ\psi$, its dual endomorphism.

\begin{theorem}[Bridge Theorem~\cite{Weiss1974}, see also~\cite{DikranjanGiordanoBruno2012}]\label{bridge}
Let $X$ be a compact metrizable abelian group whose Pontryagin dual $\widehat X$ is a torsion group, and let $\psi$ be a continuous endomorphism of $X$. Then $\entropy(\psi)=\entropy_{alg}(\widehat\psi)$.
\end{theorem}

(The Bridge Theorem of~\cite{DikranjanGiordanoBruno2012} holds for arbitrary discrete abelian groups, with the algebraic entropy defined by taking the supremum over all finite \emph{subsets} $F$ and replacing $T_m(\varphi,E)$ by $F+\varphi(F)+\cdots+\varphi^{m-1}(F)$; on torsion groups this entropy coincides with the one defined above, and in this paper $\widehat X$ is always torsion.)

The following lemma extends to matrices and to arbitrary dimension the description of the Pontryagin dual of a one-dimensional linear CA over $\ZZ/m\ZZ$ given in~\cite{ADGT2024}.

\begin{lemma}\label{lem:dual}
The Pontryagin dual of $(\Fp^n)^\zzd$ is isomorphic to the discrete group $\LP^n$, via the pairing
\[
\langle c,a\rangle=\sum_{v\in\zzd}a_v\cdot c_v\in\Fp\cong\tfrac1p\ZZ/\ZZ\subseteq\mathbb T,
\qquad c\in(\Fp^n)^\zzd,\ a=\sum_v a_vx^v\in\LP^n,
\]
where $a_v\cdot c_v$ is the standard scalar product of $\Fp^n$. Under this isomorphism, the dual of the GCA $\glorule$ with matrix $M=\sum_uA_ux^{-u}$ is the multiplication by
\[
M^*(x)=M(x^{-1})^T=\sum_{u}A_u^T\,x^{u}.
\]
\end{lemma}
\begin{proof}
Let $\chi:(\Fp^n)^\zzd\to\mathbb T$ be a continuous character. Since $(\Fp^n)^\zzd$ has exponent $p$, $\chi$ takes values in the finite, hence discrete, subgroup $\frac1p\ZZ/\ZZ\cong\Fp$ of $\mathbb T$, so that its kernel $\chi^{-1}(0)$ is open. Therefore it contains the set of configurations vanishing on some finite set $B\subseteq\zzd$; hence $\chi(c)$ depends only on $c|_B$, and $\chi$ is induced by a character of $(\Fp^n)^B$. Every character of $\Fp^n$ with values in $\Fp$ has the form $g\mapsto a\cdot g$ for a unique $a\in\Fp^n$. Thus $\chi=\langle\cdot,a\rangle$ for a unique finitely supported $a$, and conversely every such $a$ defines a continuous character. This proves the first claim. For the second one, for every $c$ and every finitely supported $a$ we have
\[
\langle\glorule(c),a\rangle=\sum_v a_v\cdot\sum_uA_uc_{v+u}=\sum_w\Big(\sum_uA_u^Ta_{w-u}\Big)\cdot c_w,
\]
so that $\widehat\glorule(a)_w=\sum_uA_u^Ta_{w-u}$, \ie, $\sum_w\widehat\glorule(a)_wx^w=\big(\sum_uA_u^Tx^u\big)\big(\sum_va_vx^v\big)$.
\end{proof}

For an $\Fp$-vector space $V$ and an endomorphism $\varphi$ of $V$ we denote by $V_\varphi$ the $\Fp[t]$-module structure on $V$ in which $t$ acts as $\varphi$. For an $\Fp[t]$-module $W$ we let $\rk(W)=\dim_{\Fp(t)}\big(\Fp(t)\otimes_{\Fp[t]}W\big)\in\N\cup\{\infty\}$; recall that $\rk(W)$ is the supremum of the cardinalities of the $\Fp[t]$-linearly independent subsets of $W$, and that $\rk(W')\leq\rk(W)$ for every submodule $W'$ of $W$, since $\Fp(t)$ is flat over $\Fp[t]$.

\begin{lemma}\label{lem:rank}
Let $V$ be an $\Fp$-vector space and $\varphi$ an endomorphism of $V$. Then $\entropy_{alg}(\varphi)=\rk(V_\varphi)\,\log p$ (with $\infty\cdot\log p=\infty$).
\end{lemma}
\begin{proof}
Since $\Fp$ is a prime field, the finite subgroups of $V$ are exactly its finite-dimensional subspaces, and $|T_m(\varphi,E)|=p^{\dim T_m(\varphi,E)}$. Hence $\entropy_{alg}(\varphi)=\log p\cdot\mathrm{ent}(\varphi)$, where $\mathrm{ent}$ is the algebraic entropy of linear transformations defined via dimensions in~\cite[Definition~2.4]{GiordanoBrunoSalce2012} (see also~\cite[Remark~2.5]{GiordanoBrunoSalce2012}), and $\mathrm{ent}(\varphi)=\rk(V_\varphi)$ by~\cite[Corollary~4.8]{GiordanoBrunoSalce2012} (see also~\cite[Equation~(1.3) and Uniqueness Theorem]{SalceVamosVirili2013})
\end{proof}

\begin{lemma}\label{lem:presentation}
Let $\mathcal A$ be a commutative $\Fp$-algebra, let $N\in M_n(\mathcal A)$, and let $\mathcal A^n_N$ denote $\mathcal A^n$ regarded as an $\Fp[t]$-module, $t$ acting as $N$. Let $\KK=\Fp(t)$ and $\mathcal A_\KK=\KK\otimes_{\Fp}\mathcal A$. Then
\[
\KK\otimes_{\Fp[t]}\mathcal A^n_N\ \cong\ \mathcal A_\KK^n\big/(tI-N)\mathcal A_\KK^n .
\]
\end{lemma}
\begin{proof}
Consider the sequence of $\mathcal A[t]$-modules
\[
\mathcal A[t]^n\xrightarrow{\ tI-N\ }\mathcal A[t]^n\xrightarrow{\ \varepsilon\ }\mathcal A^n_N\to0,
\qquad \varepsilon\Big(\sum_k t^kb_k\Big)=\sum_kN^kb_k .
\]
Clearly $\varepsilon$ is surjective and $\varepsilon\circ(tI-N)=0$. Moreover, modulo the image of $tI-N$ we have $t^kb\equiv t^{k-1}Nb$, so that every element of $\mathcal A[t]^n$ is congruent to an element of $\mathcal A^n$ (a polynomial of degree $0$ in $t$), on which $\varepsilon$ is injective; hence the sequence is exact. Since $\mathcal A[t]=\Fp[t]\otimes_{\Fp}\mathcal A$, we have $\KK\otimes_{\Fp[t]}\mathcal A[t]=\mathcal A_\KK$, and the claim follows from the right exactness of $\KK\otimes_{\Fp[t]}-$.
\end{proof}

\begin{proof}[Proof of Theorem~\ref{formula}]
Let $\KK=\Fp(t)$ and $\SSS=\KK[x_1^{\pm1},\dots,x_D^{\pm1}]=\KK\otimes_{\Fp}\LP$. The dual group $\LP^n$ of Lemma~\ref{lem:dual} is torsion. By Lemma~\ref{lem:dual}, Theorem~\ref{bridge}, Lemma~\ref{lem:rank} and Lemma~\ref{lem:presentation} (with $\mathcal A=\LP$ and $N=M^*$) we obtain
\begin{equation}\label{eq:hQ}
\entropy(\glorule)=\log p\cdot\dim_\KK Q,
\qquad Q=\SSS^n/\Gamma\,\SSS^n,\qquad \Gamma=tI-M^*\in M_n(\SSS).
\end{equation}
Since transposition does not change determinants and $x\mapsto x^{-1}$ is a ring automorphism, $\det\Gamma=\chi_M(t)|_{x\mapsto x^{-1}}$. Write $\chi_M=\sum_{u\in\zzd}g_u(t)\,x^u$ with $g_u\in\Fp[t]$, and let $\Supp(\chi_M)=\{u:\ g_u\neq0\}$. Since $g_u(t)=\sum_{i<n}(c_i)_u\,t^i+[u=0]\,t^n$, we have
\begin{equation}\label{eq:supp}
\Supp(\chi_M)=\{0\}\cup\bigcup_{i:\,c_i\neq0}\supp(c_i).
\end{equation}
In particular $\det\Gamma\neq0$, $\Supp(\det\Gamma)=-\Supp(\chi_M)$, and $c_0,\dots,c_{n-1}\in\Fp$ if and only if $\Supp(\chi_M)=\{0\}$, \ie, if and only if $\chi_M\in\Fp[t]$.

(1) Let $D=1$. Then $\SSS=\KK[x^{\pm1}]$ is a principal ideal domain. For $0\neq q\in\SSS$ let $\wdt(q)=\max\Supp(q)-\min\Supp(q)$ be its width. We have $\dim_\KK\SSS/q\SSS=\wdt(q)$: indeed $q=\lambda x^a g$ with $\lambda\in\KK^\times$, $a\in\ZZ$ and $g\in\KK[x]$ monic of degree $\wdt(q)$ with $g(0)\neq0$, and since $x$ is invertible modulo $g$, $\SSS/q\SSS=\SSS/g\SSS\cong\KK[x]/g\KK[x]$. Moreover $\wdt(qq')=\wdt(q)+\wdt(q')$, since the extreme coefficients of a product are the products of the extreme coefficients, and the units of $\SSS$ have width $0$. By the Smith normal form, $\Gamma=P\,\mathrm{diag}(d_1,\dots,d_n)\,P'$ with $P,P'\in GL_n(\SSS)$ and $d_i\neq0$, so that $Q\cong\bigoplus_i\SSS/d_i\SSS$ and
\[
\dim_\KK Q=\sum_i\wdt(d_i)=\wdt\Big(\prod_id_i\Big)=\wdt(\det\Gamma)=\wdt(\chi_M)=E^+_M-E^-_M,
\]
where the last equality follows from Equation~\eqref{eq:supp}. By Equation~\eqref{eq:hQ}, $\entropy(\glorule)=(E^+_M-E^-_M)\log p$.

(2) Let $D\geq2$. If $\chi_M\in\Fp[t]$, then $\det\Gamma\in\Fp[t]\setminus\{0\}\subseteq\KK^\times$ is a unit of $\SSS$, $\Gamma$ is invertible, $Q=0$ and $\entropy(\glorule)=0$. Otherwise $\Supp(\det\Gamma)$ contains at least two points, so that $\det\Gamma$ is not a unit of $\SSS$ (comparing extreme monomials, the units of $\SSS$ are the elements $\lambda x^u$ with $\lambda\in\KK^\times$). Since $\SSS$ is a unique factorization domain, $\det\Gamma$ has an irreducible factor $\pi$, and $\mathfrak p=\pi\SSS$ is a prime ideal containing $\det\Gamma$.

First, the localization $Q_{\mathfrak p}=\SSS_{\mathfrak p}^n/\Gamma\SSS_{\mathfrak p}^n$ is non-zero: otherwise $\Gamma B=I$ for some $B\in M_n(\SSS_{\mathfrak p})$, and $\det\Gamma$ would be a unit of $\SSS_{\mathfrak p}$, contradicting $\det\Gamma\in\mathfrak p\SSS_{\mathfrak p}$.

Now assume that $\dim_\KK Q<\infty$. For $1\leq i\leq D$, the multiplication by $x_i$ is a $\KK$-linear endomorphism of $Q$, so by the Cayley--Hamilton theorem there is $0\neq f_i\in\KK[x_i]$ with $f_iQ=0$. We claim that $f_i\notin\mathfrak p$ for some $i$. For $q\in\SSS$ and $j\leq D$, let $\wdt_j(q)$ be the width of $q$ as a Laurent polynomial in $x_j$ with coefficients in the domain $\KK[x_h^{\pm1}:h\neq j]$; as in part~(1), $\wdt_j$ is additive. If $\pi$ divides $f_1$, then $\wdt_j(\pi)=0$ for every $j\neq1$, so that $\pi$ is, up to a unit, an element of $\KK[x_1^{\pm1}]$. If $\pi$ also divides $f_2$, then in the same way $\wdt_1(\pi)=0$, and $\pi$ would be a unit, a contradiction. Hence some $f_i$ is not in $\mathfrak p$, and it is invertible in $\SSS_{\mathfrak p}$; since $f_iQ_{\mathfrak p}=0$, we get $Q_{\mathfrak p}=0$, a contradiction. Therefore $\dim_\KK Q=\infty$, and $\entropy(\glorule)=+\infty$ by Equation~\eqref{eq:hQ}.

(3) If $c_0,\dots,c_{n-1}\in\Fp$, by the Cayley--Hamilton theorem $M^n$ is an $\Fp$-linear combination of $I,M,\dots,M^{n-1}$; hence every power of $M$ belongs to the finite set $\mathrm{span}_{\Fp}\{I,M,\dots,M^{n-1}\}$, and there exist $a<b$ with $M^a=M^b$. Then $\glorule^a=\glorule^b$, the family $\{\glorule^k\}_{k\geq0}=\{\glorule^0,\dots,\glorule^{b-1}\}$ is a finite family of uniformly continuous maps, and $\glorule$ is equicontinuous. Conversely, if $\glorule$ is equicontinuous, then $\entropy(\glorule)=0$ by Proposition~\ref{prop:equi}. If $D\geq2$, (2) gives $c_0,\dots,c_{n-1}\in\Fp$. If $D=1$, (1) gives $E^+_M=E^-_M=0$, \ie, $\supp(c_i)\subseteq\{0\}$ for every $c_i\neq0$, which again means $c_0,\dots,c_{n-1}\in\Fp$.
\end{proof}

\begin{corollary}\label{cor:linearD}
Let $D\geq2$ and let $\glorule$ be a $D$-GCA on $(\ZZ/p\ZZ)^n$. If $\glorule$ is not equicontinuous (in particular, if $\glorule$ is sensitive to initial conditions), then $\entropy(\glorule)=+\infty$.
\end{corollary}

\begin{remark}\label{rem:DMM}
For $n=1$ and $D=1$ we have $M=m(x)$ and $\chi_M(t)=t-m(x)$, so that Theorem~\ref{formula} gives $\entropy(\glorule)=\big(\max(0,\max\supp(m))-\min(0,\min\supp(m))\big)\log p$, which is the formula of~\cite{DamicoMM03} for linear CAs over $\ZZ/p\ZZ$. Theorem~\ref{formula}(1) shows that, for $n>1$, the relevant quantity is not the degree range of the entries of $M$ but the one of the coefficients of its characteristic polynomial. For instance, if $M=\begin{pmatrix}0&x\\0&0\end{pmatrix}$, then $\chi_M=t^2$ and $\entropy(\glorule)=0$; indeed $\glorule^2$ is the null map.
\end{remark}

\begin{remark}\label{rem:secondproof}
Theorem~\ref{formula} also yields a second proof of Theorem~\ref{computable}, which does not rely on~\cite{BeaurK24} and gives an explicit procedure. Let $\glorule$ be a one-dimensional GCA on a finite group $G$, and let $\{(\glorule_1,G_1),\dots,(\glorule_k,G_k)\}$ be the output of the function {\tt Decomposition} of Subsection~\ref{extension} on $(\glorule,G)$; this output is effectively computable from the local rule of $\glorule$. By Corollary~\ref{cor:decomposition-D}(2), $\entropy(\glorule)=\sum_i\entropy(\glorule_i)$. If $G_i\cong(\ZZ/p\ZZ)^n$, then $\entropy(\glorule_i)$ is given by Theorem~\ref{formula}(1). Otherwise, by Lemmata~\ref{lem:terminal} and~\ref{lem:NASP_are_minimal}, either $\glorule_i$ is constant, and $\entropy(\glorule_i)=0$, or $\glorule_i$ is surjective and $G_i$ is minimal with respect to $\glorule_i$, and $\entropy(\glorule_i)$ is given by Theorem~\ref{simple_iso_copies_}(2). All these steps are effective.
\end{remark}


\section{Topological entropy of $D$-dimensional GCAs}\label{sec:Ddim}

Throughout this section, $D$ denotes an integer greater than $1$, except in Subsection~\ref{nonab}, where $D\geq1$ is arbitrary.

\medskip \noindent
We prove that  sensitive to initial conditions $D$-GCAs have infinite topological entropy. On the other hand, every $D$-GCA is either equicontinuous or sensitive to initial conditions (Proposition~\ref{prop:GCAequi}), and equicontinuous $D$-GCAs have zero topological entropy (Proposition~\ref{prop:equi}). 
As a consequence of these results, we obtain the following dichotomy.

\begin{theorem}\label{thm:dichotomy}
Let $D\geq 2$ and let $\glorule$ be a $D$-GCA on a finite group $G$. Then either $\glorule$ is equicontinuous and $\entropy(\glorule)=0$, or $\glorule$ is sensitive to initial conditions and $\entropy(\glorule)=+\infty$.
\end{theorem}

Our approach is inspired by the one-dimensional case. In \cite{CASTRONUOVO2026103749}, three of the present authors showed that the topological entropy of a one-dimensional GCA decomposes as the sum of the topological entropies of finitely many one-dimensional GCAs. More precisely, these factors are defined either on $(\ZZ/p\ZZ)^n$, with $p$ prime, or on products of isomorphic non-abelian simple groups. 

In the same paper, they also proved that a one-dimensional GCA is sensitive to initial conditions if and only if at least one of the factors in this decomposition is sensitive to initial conditions.

We prove the $D$-dimensional versions of these statements in Subsection~\ref{extension} (Corollary~\ref{cor:decomposition-D}) for a refined decomposition based on normal $\glorule$-invariant subgroups.
In particular, if a $D$-GCA is sensitive to initial conditions, then at least one of the factors arising from the corresponding decomposition is sensitive to initial conditions. Hence, to establish our main result it suffices to show that every sensitive $D$-GCA defined either on $(\ZZ/p\ZZ)^n$ (see Subsection \ref{ab}) or on a product of isomorphic non-abelian simple groups (see Subsection \ref{nonab}) has infinite topological entropy. This will imply that any $D$-GCA sensitive to initial conditions has infinite topological entropy.

\subsection{Decomposition of $D$-GCAs}\label{extension}
In this subsection we prove the $D$-dimensional versions of two results of~\cite{CASTRONUOVO2026103749} based on a decomposition of GCAs via quotients by normal $\glorule$-invariant subgroups.
In particular, we state the results related to sensitivity to initial conditions and to topological entropy that will be used throughout this paper. 

Let $G$ be a finite group and let $\glorule$ be a $D$-GCA on $G$ with local rule $f=(h_1,\dots,h_k)$. A subgroup $H\leq G$ is \emph{$\glorule$-invariant} if $\glorule(H^\zzd)\subseteq H^\zzd$. Evaluating $\glorule$ on configurations supported on a single cell, and recalling that the vectors of $V_f$ are pairwise distinct, one sees that $H$ is $\glorule$-invariant if and only if $h_j(H)\subseteq H$ for every $j$; in particular, this property can be checked effectively.
Let $H$ be a normal $\glorule$-invariant subgroup of $G$.
For $g \in G$, we denote by $[g] := gH$ the coset of $g$ in $G/H$, and for $c \in G^{\zzd}$ we write $[c]$ for the element of $(G/H)^{\zzd}$ defined by $[c](v) := [c(v)]$. The maps $\overline{\glorule}:H^\zzd\to H^\zzd$ and $\widetilde{\glorule}:(G/H)^\zzd\to (G/H)^\zzd$ defined by
\begin{eqnarray}
&&\forall c\in H^\zzd:\ \overline{\glorule}(c):=\glorule(c) \text{ and }\label{Fbar}\\
&&\forall [c]\in (G/H)^\zzd:\ \widetilde{\glorule}([c]):=[\glorule(c)] \label{Ftilde}
\end{eqnarray}
are well-defined $D$-GCAs on $H^\zzd$ and $(G/H)^\zzd$, respectively. We say that $(\glorule,G)$ is \emph{irreducible} if $G$ has no non-trivial proper normal $\glorule$-invariant subgroup.

We use the following variant of the decomposition function of~\cite{CASTRONUOVO2026103749}, where $H$ was required to be fully invariant.

\vspace{0.3cm}
\begin{algorithm}[H]
\SetAlgoNlRelativeSize{0}
\SetAlgoNoLine
\SetKwFunction{Decomposition}{Decomposition}
\SetKwProg{Fn}{Function}{:}{}
\Fn{\Decomposition{$\glorule$,$G$}}
{
\If{$(\glorule,G)$ is irreducible}
{\KwRet{$\{(\glorule,G)\}$\;}}
\Else{
Let $H$ be a non-trivial proper normal $\glorule$-invariant subgroup of $G$\;
a  $\gets$ \Decomposition{$\widetilde{\glorule},G/H$}\;
b  $\gets$ \Decomposition{$\overline{\glorule},H$}\;
\KwRet{$a\cup b$}\;}
}
\end{algorithm}
\vspace{0.3cm}
The output $\{(\glorule_1,G_1),\dots,(\glorule_k,G_k)\}$ is effectively computable from the local rule of $\glorule$, since $G$ has finitely many subgroups. Recall that a subgroup of $G$ is \emph{fully invariant} if it is mapped into itself by every endomorphism of $G$, and that $G$ is \emph{invariantly simple} if its only fully invariant subgroups are $\{e\}$ and $G$. Every fully invariant subgroup of $G$ is normal and $\glorule$-invariant; hence, if $(\glorule_i,G_i)$ is irreducible, then $G_i$ has no non-trivial proper fully invariant subgroup, \ie, it is invariantly simple, and therefore it is a direct product of isomorphic simple groups~\cite{CASTRONUOVO2026103749}. In particular, if $G_i$ is abelian, then $G_i\cong(\ZZ/p\ZZ)^n$ for some prime $p$ and some $n\in\N$.

We now prove the two facts about {\tt Decomposition} that will be used in this paper. For one-dimensional GCAs, the fact that sensitivity to initial conditions is detected by the components of the decomposition was proved in~\cite{CASTRONUOVO2026103749}. Here we give a proof which is valid in every dimension $D\geq1.$
\begin{lemma}\label{lem:sensitivity-decomposition-D}
Let $G$ be a finite group, let $\glorule$ be a $D$-GCA on $G$, and let $H$ be a normal subgroup of $G$ such that $\glorule(H^{\zzd})\subseteq H^{\zzd}$. Let $\overline{\glorule}$ be the restriction of $\glorule$ to $H^{\zzd}$ and $\widetilde{\glorule}$ the induced $D$-GCA on $(G/H)^{\zzd}$, $\widetilde\glorule([c])=[\glorule(c)]$. Then $\glorule$ is equicontinuous if and only if both $\overline{\glorule}$ and $\widetilde{\glorule}$ are equicontinuous. Consequently, $\glorule$ is sensitive to initial conditions if and only if at least one of $\overline{\glorule}$ and $\widetilde{\glorule}$ is sensitive to initial conditions.
\end{lemma}
\begin{proof}
The map $\widetilde\glorule$ is well defined because $\glorule(cz)=\glorule(c)\glorule(z)$ with $\glorule(z)\in H^\zzd$ for every $z\in H^\zzd$. If $\glorule$ is equicontinuous, then $\glorule^a=\glorule^b$ for some $a<b$ by Proposition~\ref{prop:GCAequi}(i), and the same identity holds for $\overline\glorule$ and $\widetilde\glorule$, which are therefore equicontinuous.
Conversely, assume that both $\overline\glorule$ and $\widetilde\glorule$ are equicontinuous. By Proposition~\ref{prop:GCAequi}(i) there are $a_1,a_2\geq0$ and $p_1,p_2\geq1$ with $\overline\glorule^{\,a_1+p_1}=\overline\glorule^{\,a_1}$ and $\widetilde\glorule^{\,a_2+p_2}=\widetilde\glorule^{\,a_2}$. If $\varphi^{a+p}=\varphi^{a}$, then $\varphi^{q+p}=\varphi^{q-a}\varphi^{a+p}=\varphi^{q}$ for every $q\geq a$, hence $\varphi^{2q}=\varphi^{q}$ whenever $p$ divides $q$. Therefore, choosing $q\geq\max(a_1,a_2)$ divisible by $p_1p_2$, $T=\glorule^q$ satisfies $T^2|_{H^\zzd}=T|_{H^\zzd}$ and $\widetilde T^{\,2}=\widetilde T$, where $\widetilde T=\widetilde\glorule^{\,q}$. For $x\in G^\zzd$ put $u=T(x)$ and $h=T(u)u^{-1}$. Since $[T(u)]=\widetilde T^{\,2}[x]=\widetilde T[x]=[u]$ and $H$ is normal, $h\in H^\zzd$, so that $T^2(h)=T(h)$. From $T(u)=hu$ we get, by induction on $j$,
\[
T^{j+1}(u)=T(h)^j\,h\,u\qquad(j\geq0).
\]
Let $m$ be the exponent of $H$. Since $T(h)\in H^\zzd$, we have $T(h)^m=e^\zzd$, hence $T^{m+1}(u)=T(u)$, that is, $T^{m+2}(x)=T^2(x)$. As $x$ is arbitrary, $\glorule^{q(m+2)}=\glorule^{2q}$, and $\glorule$ is equicontinuous by Proposition~\ref{prop:GCAequi}(i). The statement on sensitivity follows from Proposition~\ref{prop:GCAequi}(ii), applied to $\glorule$, $\overline\glorule$ and $\widetilde\glorule$.
\end{proof}

\begin{corollary}\label{cor:decomposition-D}
Let $(\glorule,G)$ be a $D$-GCA and let $\{(\glorule_1,G_1),\dots,(\glorule_k,G_k)\}$ be the output of {\tt Decomposition} on $(\glorule,G)$. Then:
\begin{itemize}
\item[(1)] $\glorule$ is sensitive to initial conditions if and only if some $\glorule_i$ is sensitive to initial conditions;
\item[(2)] $\entropy(\glorule)=\sum_{i=1}^k\entropy(\glorule_i)$.
\end{itemize}
\end{corollary}
\begin{proof}
Both statements are proved by induction on the depth of the recursion of {\tt Decomposition}. At each step the chosen subgroup $H$ is normal and $\glorule$-invariant. Then (1) follows from Lemma~\ref{lem:sensitivity-decomposition-D}, and (2) from the addition formula for the topological entropy of continuous endomorphisms of compact metrizable groups with respect to closed invariant normal subgroups (see, e.g.,~\cite[Theorem~19]{Bowen1971}, \cite{Yuzvinskii1965} and~\cite[Fact~3.3]{DikranjanSanchis2019}).
\end{proof}





\subsection{$D$-GCAs on $(\ZZ/p\ZZ)^n$}\label{ab}
Let $p$ be a prime. Since every endomorphism of $(\ZZ/p\ZZ)^n$ is $\Fp$-linear, $D$-GCAs on $(\ZZ/p\ZZ)^n$ are exactly the linear cellular automata studied in Section~\ref{sec:linear}; they are represented by $n \times n$ matrices whose entries are multivariate Laurent polynomials with coefficients in $\ZZ/p\ZZ$, \ie, elements of $\LP=(\ZZ/p\ZZ)[x_1,x_1^{-1},\dots,x_D,x_D^{-1}]$.

By Theorem~\ref{formula}, a $D$-GCA $\glorule$ on $(\ZZ/p\ZZ)^n$, $D\geq 2$, either is equicontinuous and has zero topological entropy (when all the coefficients of the characteristic polynomial of its matrix are constants), or is not equicontinuous and has infinite topological entropy. Since a sensitive CA is not equicontinuous, we obtain the result needed for Theorem~\ref{thm:dichotomy} (Corollary~\ref{cor:linearD}): every $D$-GCA on $(\ZZ/p\ZZ)^n$ which is sensitive to initial conditions has infinite topological entropy.

A proof of the same fact which does not use Pontryagin duality is given in~\ref{app:noduality}. We stress that neither argument requires any reduction to the Frobenius normal form of the matrix~\cite{DennunzioFormentiMargara2023}, nor any characterization of sensitivity: the dichotomy between zero and infinite entropy is a purely algebraic consequence of the fact that, for $D\geq2$, the module $\SSS^n/(tI-M^*)\SSS^n$ appearing in the proof of Theorem~\ref{formula} is either trivial or infinite-dimensional.

\subsection{$D$-GCAs on a product of isomorphic non-abelian simple groups}\label{nonab}

Throughout this subsection the dimension $D\geq1$ is arbitrary. Several of the facts collected below were established in~\cite{CASTRONUOVO2026103749} for $D=1$. Their proofs only involve the local rule and the algebraic structure of the alphabet, so they hold verbatim in every dimension; we point out explicitly the few places where the lattice $\zzd$ plays a role.

Fix a finite non-abelian simple group $S$ and let $G=S_1\times\cdots\times S_m$, with $S_i\cong S$ for every $i$. We regard each $S_i$ as a subgroup of $G$, namely as the set of elements of $G$ whose coordinates other than the $i$-th one are trivial.

Every normal subgroup of a finite direct product of non-abelian simple groups is the product of some of the factors (see, \eg,~\cite[p.~174]{dummit2003abstract}). Applied to kernels, this shows that for every $h\in\End(G)$ there is a set $I\subseteq\{1,\ldots,m\}$, possibly empty, such that $\Ker(h)=\prod_{t\in I}S_t$. Likewise, if $H=\prod_{t\in I}S_t$ is a normal subgroup of $G$, then $G/H\cong\prod_{t\notin I}S_t$ is again a product of copies of $S$; in particular, its center is trivial.

Given $v\in\zzd$ and $a\in G$, we write $c^{(v,a)}$ for the configuration which equals $a$ at position $v$ and $e$ at every other position.

\begin{lemma}\label{lem:kernel}
Let $\glorule$ be a $D$-GCA over $G = S_1 \times \cdots \times S_m$ whose local rule $f=(h_1,\ldots,h_k)$ has neighborhood $V_f=\{v_1,\ldots,v_k\}$, so that $\glorule(c)_w=h_1(c_{w+v_1})\cdots h_k(c_{w+v_k})$. Then
\[
\Ker(\glorule) = \left(\bigcap_{j=1}^k \Ker(h_j)\right)^{\mathbb Z^D}.
\]
\end{lemma}

\proof
Set $H=\bigcap_{j=1}^k\Ker(h_j)$, a normal subgroup of $G$. Every configuration with values in $H$ is mapped to $e^\zzd$, hence $H^\zzd\subseteq\Ker(\glorule)$.

For the reverse inclusion, we first determine which configurations supported on a single cell belong to the kernel. Let $v\in\zzd$ and $a\in G$. Since the vectors $v_j$ are pairwise distinct, the value of $\glorule(c^{(v,a)})$ at position $v-v_j$ is $h_j(a)$. It follows that $c^{(v,a)}\in\Ker(\glorule)$ exactly when $a\in H$.

Now take $c\in\Ker(\glorule)$, a position $v\in\zzd$ and an element $g\in G$, and put $d=c^{(v,g)}$. The kernel is a normal subgroup of $G^\zzd$, so the commutator $dcd^{-1}c^{-1}$ belongs to $\Ker(\glorule)$ as well. This commutator is trivial outside $v$, and at $v$ it equals $gc_vg^{-1}c_v^{-1}$. By the previous paragraph, $gc_vg^{-1}c_v^{-1}\in H$ for every $g\in G$; in other words, the coset $c_vH$ is central in $G/H$. As $G/H$ is centerless, $c_v\in H$, and since $v$ was arbitrary we conclude that $c\in H^\zzd$.
\endproof

\noindent The argument above works cell by cell, and therefore it does not depend on the dimension.


Recall that, by the Garden of Eden theorem of Moore and Myhill, a CA on $\zzd$ is surjective if and only if it is pre-injective. Together with Lemma~\ref{lem:kernel}, this yields the following characterization.

\begin{lemma}\label{lem:surjinj}
Let $\glorule$ be a $D$-GCA over a product of isomorphic non-abelian simple groups, with local rule $f=(h_1,\ldots,h_k)$. Then $\glorule$ is surjective if and only if it is injective, if and only if $\Ker(\glorule)=(\cap_{j\in\{1,\ldots,k\}}\Ker(h_j))^{\mathbb Z^D}$ is trivial.
\end{lemma}
\begin{proof}
Let $H=\bigcap_j\Ker(h_j)$. Assume first that $H\neq\{e\}$ and pick $a\in H\setminus\{e\}$. The finite configurations $c^{(0,a)}$ and $e^\zzd$ are distinct and have the same image; thus $\glorule$ is not injective and not even pre-injective, so that, by the Garden of Eden theorem, it is not surjective. Assume now that $H=\{e\}$. By Lemma~\ref{lem:kernel} the kernel of $\glorule$ is trivial, \ie, $\glorule$ is injective; in particular it is pre-injective, and hence surjective by the Garden of Eden theorem.
\end{proof}

\begin{lemma}\label{lem:terminal}
Let $(\glorule,G)$ be irreducible, with $G$ non-abelian. Then either $\glorule$ is the constant map $c\mapsto e^\zzd$, or $\glorule$ is surjective.
\end{lemma}
\begin{proof}
Denote by $f=(h_1,\dots,h_k)$ the local rule of $\glorule.$
By Subsection~\ref{extension}, $G$ is a product of isomorphic non-abelian simple groups. The subgroup $H=\bigcap_j\Ker(h_j)$ is normal and $\glorule(H^\zzd)=\{e^\zzd\}$, so $H$ is $\glorule$-invariant and therefore $H=G$ or $H=\{e\}$. If $H=G$, all the $h_j$ are trivial and $\glorule$ is constant. If $H=\{e\}$, then $\glorule$ is surjective by Lemma~\ref{lem:surjinj}. 
\end{proof}

From now on, $\glorule$ denotes a \textit{surjective} $D$-GCA over $G=S_1\times\cdots\times S_m$ with local rule $\locrule=(h_1,\ldots,h_k)$. Then $\locrule$ is surjective as well, since every element of $G$ is the value at the origin of some configuration in the image of $\glorule$. As shown in~\cite{CASTRONUOVO2026103749}, the endomorphisms $h_i$ then have the following structure, which only depends on $\locrule$ and not on the dimension.

\medskip\noindent
(a) The images of the $h_i$ are products of pairwise disjoint families of factors: there is a partition $J_1\cup\cdots\cup J_k$ of $\{1,\ldots,m\}$ such that $\Imma(h_i)=\prod_{t\in J_i}S_t$ for every $i$.

\medskip\noindent
(b) The kernels are products of factors as well: for every $i$ there is a set $I_i\subseteq\{1,\ldots,m\}$ with $\Ker(h_i)=\prod_{t\in I_i}S_t$ and $|I_i|=m-r_i$, where $r_i:=|J_i|$.

\medskip\noindent
(c) For every $t\in\{1,\ldots,m\}$ there is exactly one index $i$ such that $S_t$ is not a factor of $\Ker(h_i)$.

\medskip\noindent
Consequently, every $h_i$ restricts to an isomorphism from $\prod_{t\notin I_i}S_t$ onto $\prod_{t\in J_i}S_t$, both groups being isomorphic to $S^{r_i}$.

\begin{definition}\label{pifo}
Let $\pi_{\locrule}$ be the permutation of the set of factors $\{S_1,\ldots,S_m\}$ (equivalently, of the indices $\{1,\ldots,m\}$) induced by the family of endomorphisms $h_1,\ldots,h_k$: the factor $S_t$ is sent to the factor onto which it is mapped by the unique $h_i$ acting non-trivially on it. We denote by $o$ the \emph{order} of $\pi_{\locrule}$.
\end{definition}

Fix $i\in\{1,\ldots,m\}$ and follow the factor $S_i$ along $\pi_{\locrule}$. There is a unique $h_j$ acting non-trivially on $S_i$, and it maps $S_i$ onto a factor $S_{i^{(1)}}$; there is a unique $h_{j^{(1)}}$ acting non-trivially on $S_{i^{(1)}}$, and it maps $S_{i^{(1)}}$ onto a factor $S_{i^{(2)}}$; and so on. After $o$ steps we are back to $S_i$, the last map being the restriction of $h_{j^{(o-1)}}$ to $S_{i^{(o-1)}}$.
\begin{definition}\label{oi}
The automorphism $\hat h_i$ of $S_i$ is the composition
\[\hat h_i:=h_{j^{(o-1)}} \circ \ldots\circ h_{j^{(1)}}\circ h_j\enspace,
\] 
and $o_i$ denotes its order in $\Aut(S_i)$, \ie, the least positive integer such that $\hat h_i$ composed with itself $o_i$ times is the identity of $S_i$.
\end{definition}

\begin{definition}
Let $G=S_1\times\cdots\times S_m$ be a product of finite, non-abelian, isomorphic simple groups, and let $\glorule$ be a $D$-GCA over $G$. We say that $G$ is \emph{minimal} with respect to $\glorule$ if $\{1,2,\ldots,m\}$ cannot be split into two non-empty disjoint sets $I$ and $J$ such that both $\glorule(\left(\prod_{i\in I}S_i\right)^{\zzd})\subseteq \left(\prod_{i\in I}S_i\right)^{\zzd}$ and $\glorule(\left(\prod_{i\in J}S_i\right)^{\zzd})\subseteq \left(\prod_{i\in J}S_i\right)^{\zzd}$.
\end{definition}

For a surjective $\glorule$, minimality of $G$ with respect to $\glorule$ amounts to $\pi_{\locrule}$ being a single cycle.
\begin{lemma}\label{lem:NASP_are_minimal}
Let $(\glorule,G)$ be irreducible and let $\glorule$ be surjective. Then $G$ is minimal with respect to $\glorule.$
\end{lemma}
\proof
Using the same notations as above,
every sub-product $\prod_{i\in I}S_i$ is a normal subgroup of $G$; by Lemma \ref{lem:terminal} it is $\glorule$-invariant only if $I=\emptyset$ or $I=\{1,\dots,m\}$, and $G$ is minimal with respect to $\glorule$. Since $\glorule$ is assumed to be surjective, only the second case is possible. 
\endproof

Recall from Section~\ref{CAandGCA} that, for $w\in\zzd$, $\sigma^w$ denotes the shift $\sigma^w(c)_v=c_{v+w}$.

\begin{lemma}\label{lemma_product_simples}
Let $G=S_1\times\cdots\times S_m$ be a product of finite, non-abelian, isomorphic simple groups, and let $\glorule$ be a surjective $D$-GCA on $G$, with respect to which $G$ is minimal, with local rule $\locrule=(h_1,\ldots,h_k)$ and neighborhood $\{v_1,\ldots,v_k\}$. With $o$, $r_1,\ldots,r_k$ and $o_1,\ldots,o_m$ as above, we have
$$\glorule^{o\lcm(o_1,\ldots,o_m)}=\sigma^{\lcm(o_1,\ldots,o_m){\sum_i r_i v_i}}.$$  
\end{lemma}

The proof of Lemma 9 in~\cite{CASTRONUOVO2026103749} applies without changes; we sketch it to make clear that the dimension plays no role. Since $\pi_{\locrule}$ is a single cycle, $\glorule^o$ maps $S_t^\zzd$ into itself for every $t$: the $S_t$-component of $\glorule^o(c)_v$ is obtained by composing the restrictions of the $h_i$ along the cycle, and every $h_i$ occurs exactly $r_i$ times in this composition, since it acts non-trivially on exactly $r_i$ factors. Hence the $S_t$-component of $\glorule^o(c)_v$ is $\hat h_t$ applied to the $S_t$-component of $c_{v+\beta}$, where $\beta=\sum_ir_iv_i\in\zzd$ does not depend on $t$. Iterating $\lcm(o_1,\ldots,o_m)$ times turns every $\hat h_t$ into the identity, and the claim follows.

\begin{lemma}\label{lem:shiftinf}
Let $D\geq2$, let $A$ be a finite alphabet with $|A|\geq2$ and let $0\neq w\in\zzd$. Then the shift $\sigma^w$ on $A^\zzd$ has infinite topological entropy.
\end{lemma}
\begin{proof}
Let $w_1\neq0$, say (the other cases are analogous), and let $B_n=[-n,n]^D$. The sequence $(\sigma^{jw}(c)|_{B_n})_{0\leq j\leq t}$ is the restriction of $c$ to $U_t=\bigcup_{j=0}^{t}(B_n+jw)$; hence $N_{\sigma^w}(n,t)=|A|^{|U_t|}$ in the notation of Proposition~\ref{prop:Ddim}. Let $F_n=\{z\in B_n:\ z_1=n\}$, a face of $B_n$ with $(2n+1)^{D-1}$ elements. The points $z+jw$, $z\in F_n$, $0\leq j\leq t$, belong to $U_t$ and are pairwise distinct: if $z+jw=z'+j'w$ with $z,z'\in F_n$, comparing the first coordinates gives $(j-j')w_1=0$, hence $j=j'$ and $z=z'$. Therefore $|U_t|\geq(t+1)(2n+1)^{D-1}$, and $\lim_t\frac1t\log N_{\sigma^w}(n,t)\geq(2n+1)^{D-1}\log|A|$. By Proposition~\ref{prop:Ddim}, $\entropy(\sigma^w)=+\infty$.
\end{proof}

Lemma~\ref{lemma_product_simples} has the following consequences.

\begin{theorem}\label{simple_iso_copies_}
Let $G=S_1\times\cdots\times S_m$ with $S_i\cong S$ for every $i$, where $S$ is a finite non-abelian simple group, and let $\glorule$ be a surjective $D$-GCA on $G$, with respect to which $G$ is minimal, with local rule $\locrule=(h_1,\ldots,h_k)$ and neighborhood $\{v_1,\ldots,v_k\}$. Let $r_1,\ldots,r_k$ be as in Lemma~\ref{lemma_product_simples} and $\beta=\sum_ir_iv_i$. Then:\\
$(1)$ the following are equivalent: $\glorule$ is sensitive to initial conditions; $\glorule$ is topologically transitive; $\beta\neq0$;\\
$(2)$ the topological entropy of $\glorule$ is
$$\entropy(\glorule)=\begin{cases}
    |\beta|\log|S| & \mbox{ if }D=1,\\
    +\infty & \mbox{ if }D>1\mbox{ and }\beta\neq0,\\
    0 & \mbox{ if }D>1\mbox{ and }\beta=0.
\end{cases}$$
\end{theorem}
\begin{proof}
Let $L=\lcm(o_1,\ldots,o_m)$, $N=oL$ and $s=L\beta$, so that $\glorule^N=\sigma^s$ by Lemma~\ref{lemma_product_simples}.

$(1)$ If $\beta\neq0$, then $s\neq0$, and the shift $\sigma^s$ on $G^\zzd$ is sensitive to initial conditions and topologically transitive (indeed mixing). Since the iterates of $\glorule^N$ are iterates of $\glorule$, sensitivity and transitivity of $\glorule^N$ imply those of $\glorule$. If $\beta=0$, then $\glorule^N$ is the identity; hence $\glorule$ is equicontinuous by Proposition~\ref{prop:GCAequi}(i), so it is not sensitive, and it is not transitive: given $x$ and $y$ outside the finite orbit $\{x,\glorule(x),\dots,\glorule^{N-1}(x)\}$, by continuity there are open sets $U\ni x$ and $V\ni y$ with $\glorule^j(U)\cap V=\emptyset$ for $0\leq j<N$, hence for every $j\geq0$.

$(2)$ If $\beta=0$, then $\glorule^N$ is the identity and $\entropy(\glorule)=0$. Let $\beta\neq0$. By fact (a) of Section~\ref{sec:entropy}, $\entropy(\glorule)=\entropy(\sigma^s)/N$. If $D>1$, then $\entropy(\sigma^s)=+\infty$ by Lemma~\ref{lem:shiftinf}. If $D=1$, then $\entropy(\sigma^s)=|s|\log|G|$, whence
\[
\entropy(\glorule)=\frac{|s|\log|G|}{N}=\frac{|\beta|\log|G|}{o}=|\beta|\log|S|,
\]
since $o=m$, the permutation $\pi_\locrule$ being a single cycle on the $m$ factors, and $|G|=|S|^m$. Note that the value of $L$ plays no role.
\end{proof}

\subsection{Proof of Theorem~\ref{thm:dichotomy}}

Let $\glorule$ be a $D$-GCA, $D\geq2$. If $\glorule$ is equicontinuous, then $\entropy(\glorule)=0$ by Proposition~\ref{prop:equi}. Otherwise $\glorule$ is sensitive to initial conditions by Proposition~\ref{prop:GCAequi}. Let $\{(\glorule_1,G_1),\dots,(\glorule_k,G_k)\}$ be the output of {\tt Decomposition} on $(\glorule,G)$. By Corollary~\ref{cor:decomposition-D}(1), some $\glorule_i$ is sensitive, and by Corollary~\ref{cor:decomposition-D}(2) it suffices to prove that $\entropy(\glorule_i)=+\infty$. If $G_i\cong(\ZZ/p\ZZ)^n$, this follows from Corollary~\ref{cor:linearD}. Otherwise, by Lemma~\ref{lem:terminal}, $\glorule_i$ is not constant, since it is sensitive; hence it is surjective and $G_i$ is minimal with respect to $\glorule_i$, and $\entropy(\glorule_i)=+\infty$ by Theorem~\ref{simple_iso_copies_}. \qed

\begin{corollary}\label{cor:computableD}
For every $D\geq1$ there is an algorithm which, given a finite group $G$ by its multiplication table and the local rule of a $D$-GCA $\glorule$ on $G$, computes $\entropy(\glorule)\in[0,+\infty]$.
\end{corollary}
\begin{proof}
For $D=1$ this is Theorem~\ref{computable}. For $D\geq2$, by Theorem~\ref{thm:dichotomy} we have $\entropy(\glorule)=0$ if $\glorule$ is equicontinuous and $\entropy(\glorule)=+\infty$ otherwise, and equicontinuity of group cellular automata is decidable in every dimension~\cite[Theorem~27]{BeaurK24}.
\end{proof}

\noindent For $D\geq2$, the decomposition of Subsection~\ref{extension} also gives a direct decision procedure, which does not rely on~\cite{BeaurK24}: $\entropy(\glorule)=+\infty$ if and only if some irreducible component of $\glorule$ is either abelian with a non-constant coefficient in its characteristic polynomial, or non-abelian, non-constant and with $\beta\neq0$ (see Section~\ref{sec:complexity}).

\section{Computational complexity}\label{sec:complexity}

In this section we make the procedures of Remark~\ref{rem:secondproof} and Corollary~\ref{cor:computableD} explicit and bound their cost. 

\medskip\noindent
\textbf{Input.}
The group $G$, of order $N$, is given by its multiplication table, and the local rule $f=(h_1,\dots,h_k)$ of $\glorule$ by the $k$ lookup tables of the endomorphisms $h_j$ together with the neighborhood $V_f=\{v_1,\dots,v_k\}\subseteq\zzd$. The size of the input is $\Theta(N^2+kN+kD)$. We let $\varrho=\max_j\|v_j\|_\infty$, and we assume that the group operation, the maps $h_j$, arithmetic in $\Fp$ and additions of integers cost $O(1)$.

\medskip\noindent
\textbf{The algorithm.}
The procedure {\tt Entropy} first calls {\tt Decomposition} (Subsection~\ref{extension}), and then processes each irreducible pair $(\glorule_i,G_i)$ of its output:
\begin{itemize}
\item[(A)] if $G_i$ is abelian, it computes an isomorphism $G_i\cong\Fp^n$, the matrix $M_i=M_{\glorule_i}$ and its characteristic polynomial $\chi_{M_i}$; for $D=1$ it records the pair $(p,\,E^+_{M_i}-E^-_{M_i})$, and for $D\geq2$ it marks $\glorule_i$ as sensitive if some coefficient of $\chi_{M_i}$ is not constant;
\item[(N)] if $G_i$ is non-abelian, it checks whether all the endomorphisms of the local rule of $\glorule_i$ are trivial; if not, it computes $|S|$, where $G_i\cong S^m$, the integers $r_j$ defined by $|\Imma(h_j)|=|S|^{r_j}$, and $\beta=\sum_jr_jv_j$; for $D=1$ it records the pair $(|S|,|\beta|)$, and for $D\geq2$ it marks $\glorule_i$ as sensitive if $\beta\neq0$.
\end{itemize}
For $D=1$ the output is $\entropy(\glorule)=\sum a\log b$, the sum ranging over the recorded pairs $(b,a)$. For $D\geq2$ the output is $+\infty$ if some $\glorule_i$ is marked, and $0$ otherwise.

\medskip\noindent
\textbf{Correctness.}
Let $D=1$. By Corollary~\ref{cor:decomposition-D}(2), $\entropy(\glorule)=\sum_i\entropy(\glorule_i)$. In case (A), $\entropy(\glorule_i)=(E^+_{M_i}-E^-_{M_i})\log p$ by Theorem~\ref{formula}(1). In case (N), by Lemmata~\ref{lem:terminal} and~\ref{lem:NASP_are_minimal}  either $\glorule_i$ is constant, and $\entropy(\glorule_i)=0$, or $\glorule_i$ is surjective and $G_i$ is minimal with respect to $\glorule_i$. In the latter case $\Imma(h_j)=\prod_{t\in J_j}S_t$ by fact (a) of Subsection~\ref{nonab}, so that $r_j=|J_j|$, and $\entropy(\glorule_i)=|\beta|\log|S|$ by Theorem~\ref{simple_iso_copies_}(2).

Let $D\geq2$. By Theorem~\ref{thm:dichotomy}, $\entropy(\glorule)=+\infty$ if $\glorule$ is sensitive and $\entropy(\glorule)=0$ otherwise. By Corollary~\ref{cor:decomposition-D}(1), $\glorule$ is sensitive if and only if some $\glorule_i$ is. In case (A), $\glorule_i$ is sensitive if and only if some coefficient of $\chi_{M_i}$ is not constant, by Theorem~\ref{formula}(3) and Proposition~\ref{prop:GCAequi}(ii). In case (N), a constant $\glorule_i$ is equicontinuous; otherwise, by Lemma~\ref{lem:terminal} and Theorem~\ref{simple_iso_copies_}(1), $\glorule_i$ is sensitive if and only if $\beta\neq0$.

\medskip\noindent
\textbf{Auxiliary results.}

\begin{lemma}\label{lem:closure}
Let $G$ be a finite group of order $M$, given by its multiplication table, and let $\varphi_1,\dots,\varphi_s$ be endomorphisms of $G$, given by their tables. One can find a normal subgroup $\{e\}\neq H\neq G$ with $\varphi_i(H)\subseteq H$ for every $i$, or decide that no such subgroup exists, in $O(M^3+sM^2)$ operations. With $s=0$, the same procedure computes the normal closures of all the elements of $G$ within the same bound.
\end{lemma}
\begin{proof}
The identity and the table of inverses are computed in $O(M^2)$ operations. For $a\in G$, let $\mathrm{Cl}(a)$ be the smallest normal subgroup of $G$ that contains $a$ and is invariant under $\varphi_1,\dots,\varphi_s$. It is computed by a breadth-first closure: starting from $H=\{a\}$, every element $x$ added to $H$ is processed once, and for each such $x$ one adds to $H$ the elements $x^{-1}$, $xy$ and $yx$ for all current $y\in H$, $g^{-1}xg$ for all $g\in G$, and $\varphi_i(x)$ for all $i$. When an element $y$ enters $H$ after $x$, the products $xy$ and $yx$ are added when $y$ is processed; hence the result is closed under products, inverses, conjugation and the maps $\varphi_i$, and it is the smallest such set containing $a$. Each element is processed at most once, at a cost of $O(M+s)$, so that one closure costs $O(M^2+sM)$. The algorithm computes $\mathrm{Cl}(a)$ for every $a\neq e$ and returns the first one different from $G$. This is correct: if $K$ is a subgroup with the required properties and $a\in K\setminus\{e\}$, then $\mathrm{Cl}(a)\subseteq K\neq G$. The total cost is $O(M^3+sM^2)$.
\end{proof}

\begin{lemma}\label{lem:coordinates}
Let $G\cong\Fp^n$ be given by its multiplication table, with $M=|G|=p^n$, and let $h_1,\dots,h_k$ be endomorphisms of $G$, given by their tables. One can compute $p$, a basis $g_1,\dots,g_n$ of $G$ over $\Fp$, a table of the coordinates of all the elements of $G$ in this basis, and the matrices $A_1,\dots,A_k\in M_n(\Fp)$ of $h_1,\dots,h_k$, in $O(M\log M+kM)$ operations.
\end{lemma}
\begin{proof}
The prime $p$ is the order of any element $g\neq e$, computed in $O(p)\subseteq O(M)$ operations. Let $V_0=\{e\}$. Given the span $V_i$ of $g_1,\dots,g_i$, together with the coordinates of its elements, choose $g_{i+1}\notin V_i$ by scanning $G$ with a membership array. Then $V_{i+1}=\bigcup_{a\in\Fp}g_{i+1}^aV_i$, and the coordinates of its elements are obtained from those of $V_i$, in $O(n|V_{i+1}|+M)$ operations. Since $|V_{i+1}|=p^{i+1}$ and $n\leq\log_2M$, the whole construction costs $O(nM)=O(M\log M)$. The $j$-th column of $A_l$ is the coordinate vector of $h_l(g_j)$, read from the tables in $O(n)$ operations. Reading the $k$ tables costs $O(kM)$, which dominates the $O(kn^2)$ operations needed to fill the matrices.
\end{proof}

\medskip\noindent
\textbf{Size of the recursion tree.}
Let $\lambda(G)$ be the composition length of $G$, \ie, the length of any composition series of $G$ (see~\cite[p.~108]{hungerford2003algebra}). By the Jordan--H\"older theorem, $\lambda(G)=\lambda(H)+\lambda(G/H)$ for every normal subgroup $H$ (see~\cite[p.~111]{hungerford2003algebra}). Every leaf of the recursion tree of {\tt Decomposition} carries a non-trivial group, which has composition length at least $1$. Hence the tree has at most $\lambda(G)\leq\log_2N$ leaves, and, every internal node having two children, at most $2\log_2N$ nodes. Moreover, since $|G|=|H|\,|G/H|$, the orders of the groups attached to the nodes at a fixed depth multiply to at most $N$, and the orders $M_1,\dots,M_q$ of the groups at the leaves multiply to exactly $N$. Since each child has order at most half the order of its parent, the depth of the tree is at most $\log_2N$.

\medskip\noindent
\textbf{Cost at a node.}
Consider a node with a group of order $M$. Deciding irreducibility, and finding $H$ if the node is internal, costs $O(M^3+kM^2)$ by Lemma~\ref{lem:closure} with $\varphi_j=h_j$; recall from Subsection~\ref{extension} that $H$ is $\glorule$-invariant if and only if $h_j(H)\subseteq H$ for every $j$. At an internal node one also builds the multiplication tables of $H$ and $G/H$, in $O(M^2)$ operations, and the tables of the restrictions $h_j|_H$ and of the induced endomorphisms $[g]\mapsto[h_j(g)]$ of $G/H$, in $O(kM)$ operations. The neighborhood does not change. Since $x\mapsto x^3$ is superadditive on positive integers, the sum of $M^3+kM^2$ over the nodes at a fixed depth is at most $N^3+kN^2$. Hence all nodes together cost
\[
O\big((N^3+kN^2)\log N\big).
\]

\medskip\noindent
\textbf{Cost at a leaf.}
Let $(\glorule_i,G_i)$ be a leaf, with $M=|G_i|$. Testing whether $G_i$ is abelian costs $O(M^2)$ operations.

\emph{Case (A).} Let $G_i\cong\Fp^n$, so that $n=\log_pM\leq\log_2M$. By Lemma~\ref{lem:coordinates}, the matrices of the $h_j$ are obtained in $O(M\log M+kM)$ operations, and $M_i=\sum_jA_jx^{-v_j}$ in $O(kn^2D)$ operations. We compute $\chi_{M_i}$ with Berkowitz's division-free algorithm~\cite{Berkowitz1984}, which uses $O(n^4)$ ring operations. All intermediate quantities are polynomials of degree at most $n$ in the entries of $M_i$. They are therefore Laurent polynomials with exponents in $[-n\varrho,n\varrho]^D$, hence with at most $(2n\varrho+1)^D$ monomials, and each ring operation costs $O\big(D(2n\varrho+1)^{2D}\big)$. Reading off $E^\pm_{M_i}$, or testing whether the coefficients are constant, is cheaper. The cost of case (A) is
\[
O\big(M\log M+kM+kn^2D+Dn^4(2n\varrho+1)^{2D}\big).
\]

\emph{Case (N).} Testing whether all the $h_j$ are trivial costs $O(kM)$. For $a\neq e$, the normal closure of $a$ in $G_i\cong S^m$ is the product of the factors in which $a$ has a non-trivial coordinate. Hence $|S|$ is the smallest order of the normal closure of an element $a\neq e$, computed in $O(M^3)$ operations by Lemma~\ref{lem:closure} with $s=0$. The orders $|\Imma(h_j)|$, and hence the $r_j$, cost $O(kM)$, and $\beta$ costs $O(kD)$. The cost of case (N) is $O(M^3+kM+kD)$.

\medskip\noindent
\textbf{Overall complexity.}
Recall that the leaf orders $M_1,\dots,M_q$ satisfy $\prod_iM_i=N$ and $M_i\geq2$; hence $\sum_iM_i^3\leq N^3$ and $\sum_iM_i\log M_i\leq N\log N$. Moreover, the exponents $n_i$ of the abelian leaves satisfy $\sum_in_i\leq\log_2N$, so that
\[
\sum_in_i^4(2n_i\varrho+1)^{2D}\leq(\log_2N)^4(2\varrho\log_2N+1)^{2D}.
\]
Adding the costs of the nodes and of the leaves, we obtain the following result.

\begin{theorem}\label{thm:complexity}
The procedure {\tt Entropy} computes the topological entropy of a $D$-GCA $\glorule$ on a group of order $N$, with $k$ neighbors and $\varrho=\max_{v\in V_f}\|v\|_\infty$, in
\[
O\Big((N^3+kN^2+kD)\log N+D\,(\log N)^4\,(2\varrho\log_2N+1)^{2D}\Big)
\]
operations. In particular, for every fixed $D$ the running time is polynomial in $N$, $k$ and $\varrho$. For $D=1$ it is
\[
O\big((N^3+kN^2)\log N+\varrho^2(\log N)^6\big).
\]
\end{theorem}

\section{Tables}

\begin{remark}\label{rem:semigroups}
Every CA is a CA over a semigroup. Indeed, endow the alphabet $A$ with the left-zero operation $a\cdot b=a$, which is associative. The induced componentwise operation on $A^\zzd$ is again the left-zero operation, so that every map $\glorule:A^\zzd\to A^\zzd$ satisfies $\glorule(x\cdot y)=\glorule(x)=\glorule(x)\cdot\glorule(y)$, \ie, it is a semigroup endomorphism of $A^\zzd$. Consequently, by~\cite{Hurd_Kari_Culik_1992}, the topological entropy of CAs over semigroups is not computable. This argument does not apply to monoids: a left-zero semigroup with at least two elements has no identity element, and the global map of a CA over a monoid must fix the identity configuration.
\end{remark}

\begin{table}[h]
    \centering
    \begin{tabular}{|c|c|c|} \hline
    $D=1$    & exact formula  & computable/uncomputable \\ \hline
        Cyclic groups & given in \cite{DamicoMM03} & \makecell{ computable \\ implied by 
        \cite{DamicoMM03} }   \\ \hline
       Abelian groups  & \makecell{explicit algorithm \\ $[$this paper$]$} & \makecell{ computable \\  
       $[$this paper$]$}  \\ \hline
       Groups  & \makecell{explicit algorithm \\ $[$this paper$]$} &  \makecell{\textbf{computable} \\  
       \textbf{$[$this paper$]$}}  \\ \hline
       Monoids  & open problem & \makecell{ conjectured to be computable \\  
       $[$an extension of this paper$]$}  \\ \hline
      \makecell{ Semigroups \\ (General CA)}  & \makecell{no general computable \\ formula, by \cite{Hurd_Kari_Culik_1992}} & \makecell{ uncomputable \\  
        \cite{Hurd_Kari_Culik_1992}} \\ \hline
    \end{tabular}
    \caption{Topological entropy of one-dimensional CAs over algebraic alphabets: exact formulas and computability.}
    \label{tab:dim1}
\end{table}

\begin{table}[h]
    \centering
    \begin{tabular}{|c|c|c|c|} \hline
    $D>1$    & $\entropy =  0 $ & $\entropy = \infty$ &  $0<\entropy<\infty $\\ \hline
        Cyclic groups & equicontinuous & sensitive \cite{DamicoMM03} & empty class \cite{DamicoMM03} \\ \hline
       Abelian groups  &  equicontinuous & \makecell{sensitive $[$this paper$]$ \\ (only for mixing \cite{MorrisWard1998})}
       & \makecell{empty class \\ $[$this paper$]$}\\ \hline
       Groups  & equicontinuous & \makecell{\textbf{sensitive} \\ \textbf{$[$this paper$]$}} &  \makecell{an empty class \\ $[$this paper$]$}\\ \hline
       Monoids  & \makecell{ conjectured to be \\ equicontinuous \\  
       $[$ext. this paper$]$} & \makecell{ conjectured to be \\ sensitive \\  
       $[$ext. this paper$]$} & \makecell{ conjectured to be \\ empty class \\  
       $[$ext. this paper$]$}\\ \hline
       \makecell{ Semigroups \\ (General CA)}  & undecidable \cite{Hurd_Kari_Culik_1992} & undecidable
       & \makecell{non-empty class \\ \cite{Meyerovitch2008}}\\ \hline
    \end{tabular}
    \caption{Topological entropy of $D$-dimensional CAs over algebraic alphabets, $D>1$.}
    \label{tab:dimD}
\end{table}

\section{Conclusions and Further Work}

In this paper we investigate the problem of computing topological entropy for
cellular automata, emphasizing the role of algebraic structure on the alphabet
and on the local rule. While topological entropy is known to be algorithmically
intractable for general cellular automata over arbitrary finite alphabets and
local rules, we show that this negative picture changes substantially when the
alphabet is endowed with a group structure and the local rule is a group
homomorphism.

Our main result proves that the topological entropy of group
cellular automata is computable in every dimension. 
This extends earlier computability results for linear cellular automata over
rings of the form $\ZZ/m\ZZ$, obtained in~\cite{DamicoMM03}, to the much broader class of cellular automata defined over arbitrary
finite groups.
The key ingredient is a structural decomposition of space-time behavior into
independent left and right contributions, which allows the entropy to be
recovered from the growth rate of fixed width space-time patterns.
As a consequence, the entropy computation can be reduced to an effective
procedure, in sharp contrast with the general incomputability results known for
unrestricted cellular automata. Moreover, via Pontryagin duality and the Bridge Theorem, we obtained an explicit formula for linear cellular automata over $(\ZZ/p\ZZ)^n$, in terms of the characteristic polynomial of the associated matrix. We also showed that the resulting procedure runs in polynomial time in the size of the group, the number of neighbors and the radius, for every fixed dimension (Theorem~\ref{thm:complexity}).

Beyond the specific result on computability, our work highlights how algebraic
constraints on the alphabet and on the local rule can dramatically simplify the dynamical complexity
of cellular automata and restore algorithmic tractability of fundamental
invariants.
This contributes to a finer understanding of the boundary between computability
and incomputability phenomena in symbolic and cellular dynamics.

Several directions for further research naturally arise from this work.
A first question concerns the extension of our techniques beyond groups.
We conjecture that the same approach can be adapted to cellular automata defined
over finite monoids, leading again to decidability results for
topological entropy.
On the other hand, for more general algebraic structures such as semigroups,
strong incomputability results are already known, suggesting that the group and
monoid cases represent a genuine frontier between tractable and intractable
settings.

A second direction concerns higher-dimensional cellular automata.
For group cellular automata in dimension $D\geq2$ we proved that either the automaton is equicontinuous and has zero topological entropy, or it is sensitive to initial conditions and has infinite topological entropy (Theorem~\ref{thm:dichotomy}).
This answers, in the setting of full shifts, the algebraic version of Shereshevsky’s question formulated in~\cite{MorrisWard1998}.
For general cellular automata no such dichotomy holds, since in every dimension $D\geq 2$ there exist cellular automata with finite positive entropy~\cite{Meyerovitch2008}. It would be interesting to understand whether the dichotomy extends to cellular automata over finite monoids, and more generally which algebraic constraints force it.

Finally, it would be interesting to investigate whether the techniques introduced
here can be applied to other quantitative invariants of cellular automata.
In particular, a natural question is whether the decomposition of space-time
behavior developed in this paper can be exploited to compute or better
understand Lyapunov exponents for group cellular automata, providing a unified
framework for the effective analysis of different indicators of dynamical
instability.

We believe that these directions offer promising avenues for further research
and that the approach developed in this work may prove useful beyond the
specific problem of topological entropy, contributing more broadly to the study
of algebraic cellular automata and their dynamical properties.

\appendix

\section{A proof of Theorem~\ref{formula}(2)--(3) without Pontryagin duality}\label{app:noduality}

In this appendix we prove parts (2) and (3) of Theorem~\ref{formula} without using Pontryagin duality and the Bridge Theorem: we only use the one-dimensional results of Section~\ref{sec:mainresult} and elementary algebra. The idea is to bound from below the entropy of a $D$-GCA by the entropy of its restriction to spatially periodic configurations. Such a restriction is a one-dimensional GCA over a larger alphabet, which splits into many independent one-dimensional linear CAs, each contributing at least $\log p$ to the entropy unless it is equicontinuous. Part (1) of Theorem~\ref{formula}, \ie, the exact value of the entropy in dimension one, is not recovered by this argument.

In this appendix the dimension $D\geq1$ is arbitrary, except in the statements explicitly concerning one-dimensional GCAs. For a finite field $\mathbb E$ of characteristic $p$ we consider linear CAs on $(\mathbb E^n)^\zzd$, defined as in Section~\ref{sec:linear} with $\mathbb E$ in place of $\Fp$: $\glorule(c)_v=\sum_{u}A_uc_{v+u}$ with $A_u\in M_n(\mathbb E)$, with associated matrix $M_\glorule=\sum_uA_ux^{-u}\in M_n(\mathbb E[x_1^{\pm1},\dots,x_D^{\pm1}])$. They are $D$-GCAs on the elementary abelian $p$-group $\mathbb E^n$. As in Section~\ref{sec:linear}, $M_{\glorule^k}=M_\glorule^k$, and $\glorule$ is determined by $M_\glorule$ and vice versa.

\begin{lemma}\label{lem:zeroequi}
Let $\glorule$ be a one-dimensional GCA on a finite group $G$. If $\entropy(\glorule)=0$, then $\glorule$ is equicontinuous.
\end{lemma}
\begin{proof}
Write $l=l_\glorule$ and $r=r_\glorule$. By Corollary~\ref{cor:logint}, $N^-_\infty$ and $N^+_\infty$ are trivial, so that, by Proposition~\ref{prop:structure}, $N^-_t$ and $N^+_t$ are trivial for every $t\geq t_0$.

\emph{Step 1.} Let $U\in L$ be such that $\glorule^j(U)$ is trivial on $[-l+1,0]$ for $0\leq j\leq t_0$. Then $\glorule^j(U)$ is trivial on $[-l+1,0]$ for every $j\geq0$: if this holds for $j\leq t$ with $t\geq t_0$, then $\glorule^{t+1}(U)_{[-l+1,0]}\in N^-_t=\{e\}$. By Lemma~\ref{lem:determinacy}(i), $\glorule^j(U)$ is trivial on $[-l+1,+\infty)$ for every $j\geq0$. Symmetrically, if $U\in R$ and $\glorule^j(U)$ is trivial on $[0,r-1]$ for $0\leq j\leq t_0$, then $\glorule^j(U)$ is trivial on $(-\infty,r-1]$ for every $j\geq0$.

\emph{Step 2.} Let $K=(t_0+1)l$ and $K'=(t_0+1)r$. Since $\glorule(c)_i$ only depends on $c_{[i-l,i+r]}$, if $U$ is trivial on $(-K,+\infty)$ then $\glorule^j(U)$ is trivial on $(-K+jl,+\infty)$, and in particular on $[-l+1,0]$ for $j\leq t_0$. By Step 1 and shift-commutation, for every $a\in\ZZ$: if $U$ is trivial on $(a-K,+\infty)$, then every $\glorule^j(U)$ is trivial on $[a-l+1,+\infty)$. Symmetrically, if $U$ is trivial on $(-\infty,a+K')$, then every $\glorule^j(U)$ is trivial on $(-\infty,a+r-1]$.

\emph{Step 3.} Let $m\geq0$ and let $c,c'\in G^\ZZ$ coincide on $[-m-K,m+K']$. Then $U=c^{-1}c'=U_LU_R$, where $U_L$ is trivial on $(-m-K-1,+\infty)$ and $U_R$ is trivial on $(-\infty,m+K'+1)$. By Step 2 with $a=-m+l-1$ (note that $-m-K-1\leq a-K$) and with $a=m+1$, every $\glorule^j(U_L)$ is trivial on $[-m,+\infty)$ and every $\glorule^j(U_R)$ is trivial on $(-\infty,m+r]$. Hence $\glorule^j(c')=\glorule^j(c)\glorule^j(U_L)\glorule^j(U_R)$ coincides with $\glorule^j(c)$ on $[-m,m]$ for every $j\geq0$, and $\glorule$ is equicontinuous.
\end{proof}

\begin{corollary}\label{cor:gap1D}
Let $\glorule$ be a one-dimensional GCA on a finite non-trivial group $G$, and let $\varpi$ be the smallest prime divisor of $|G|$. Then either $\glorule$ is equicontinuous and $\entropy(\glorule)=0$, or $\glorule$ is sensitive to initial conditions and $\entropy(\glorule)\geq\log\varpi$.
\end{corollary}
\begin{proof}
By Propositions~\ref{prop:equi} and~\ref{prop:GCAequi} and Lemma~\ref{lem:zeroequi}, $\glorule$ is equicontinuous if and only if $\entropy(\glorule)=0$, and otherwise it is sensitive. By Corollary~\ref{cor:logint}, $\entropy(\glorule)=\log k$ with $k$ dividing $|G|^{l_\glorule+r_\glorule}$; if $k>1$, then $k\geq\varpi$.
\end{proof}

\begin{lemma}\label{lem:equichar}
Let $\mathbb E$ be a finite field, $D\geq1$, and let $\glorule$ be a linear CA on $(\mathbb E^n)^\zzd$ with matrix $M$. If $\glorule$ is equicontinuous, then all the coefficients of $\chi_M(t)$ belong to $\mathbb E$.
\end{lemma}
\begin{proof}
By Proposition~\ref{prop:GCAequi}(i) there are $a<b$ with $\glorule^a=\glorule^b$, hence $M^a=M^b$. Let $\Omega$ be an algebraic closure of the field $\mathbb E(x_1,\dots,x_D)$ of rational functions. If $\lambda\in\Omega$ is a root of $\chi_M$, then $\lambda$ is an eigenvalue of $M$, so $\lambda^a=\lambda^b$; hence either $\lambda=0$ or $\lambda^{b-a}=1$, and in both cases $\lambda$ is algebraic over $\mathbb E$. The coefficients of $\chi_M$ are, up to sign, elementary symmetric functions of its roots, so they are algebraic over $\mathbb E$. Since $\mathbb E(x_1,\dots,x_D)$ is a purely transcendental extension of $\mathbb E$, the field $\mathbb E$ is relatively algebraically closed in $\mathbb E(x_1,\dots,x_D)$; as the coefficients of $\chi_M$ belong to $\mathbb E(x_1,\dots,x_D)$ and are algebraic over $\mathbb E$, they belong to $\mathbb E$. 
\end{proof}

\begin{theorem}\label{thm:noduality}
Let $D\geq2$ and let $\glorule$ be a $D$-GCA on $\Fp^n$ with matrix $M$. If some coefficient of $\chi_M$ does not belong to $\Fp$, then $\entropy(\glorule)=+\infty$.
\end{theorem}
\begin{proof}
Some coefficient $c_k$ of $\chi_M$ contains a monomial $x^u$ with $u\neq0$; after permuting the coordinates we may assume $u_1\neq0$. Write $x'=(x_2,\dots,x_D)$ and $c_k=\sum_{a\in\ZZ}x_1^a\,b_a(x')$ with $b_a\in\Fp[x_2^{\pm1},\dots,x_D^{\pm1}]$, and let $b=b_{u_1}\neq0$.

Fix $s\geq1$, let $q=p^s-1$, $\mathbb E=\mathbb F_{p^s}$, and let
\[
Y_q=\{c\in(\Fp^n)^\zzd:\ c_{v+qe_i}=c_v\ \text{for every } v\in\zzd \text{ and } 2\leq i\leq D\}.
\]
The set $Y_q$ is closed and $\glorule(Y_q)\subseteq Y_q$, so that $\entropy(\glorule)\geq\entropy(\glorule|_{Y_q})$ by fact (c) of Section~\ref{sec:entropy}.

Let $\mathcal R_q=\Fp[x_2^{\pm1},\dots,x_D^{\pm1}]/(x_2^q-1,\dots,x_D^q-1)$, the group algebra of $(\ZZ/q\ZZ)^{D-1}$. Grouping the cells of a configuration in $Y_q$ according to their first coordinate, $Y_q$ is identified with $(\mathcal R_q^n)^\ZZ$, and $\glorule|_{Y_q}$ with the one-dimensional CA which acts on finite configurations, \ie, on $\mathcal R_q[x_1^{\pm1}]^n$, as the multiplication by the image $M_q$ of $M$ in $M_n(\mathcal R_q[x_1^{\pm1}])$.

Let $Z=(\mathbb E^\times)^{D-1}$; it is the set of common zeros of $x_2^q-1,\dots,x_D^q-1$, and it has $q^{D-1}$ elements. The Frobenius map $\phi(\zeta)=(\zeta_2^p,\dots,\zeta_D^p)$ acts on $Z$; let $O_1,\dots,O_N$ be its orbits, choose $\zeta^{(j)}\in O_j$, and let $d_j=|O_j|$, so that $\Fp(\zeta^{(j)})=\mathbb F_{p^{d_j}}$ and $d_j\leq s$. Since $p\nmid q$, the polynomial $x^q-1$ has $q$ distinct roots in $\mathbb E$, and the evaluation map
\[
\mathrm{ev}:\mathcal R_q\to\prod_{j=1}^N\mathbb F_{p^{d_j}},\qquad g\mapsto\big(g(\zeta^{(j)})\big)_j,
\]
is an isomorphism of $\Fp$-algebras: it is injective, because $\mathcal R_q\otimes_{\Fp}\mathbb E\cong\mathbb E^Z$ via evaluation at all points of $Z$ and $g(\phi(\zeta))=g(\zeta)^p$, and both sides have dimension $q^{D-1}=\sum_jd_j$ over $\Fp$. Applying $\mathrm{ev}$ coefficientwise and cellwise, we obtain a group isomorphism and homeomorphism $\Psi:(\mathcal R_q^n)^\ZZ\to\prod_j(\mathbb F_{p^{d_j}}^n)^\ZZ$ which conjugates $\glorule|_{Y_q}$ to $\prod_j H_j$, where $H_j$ is the one-dimensional linear CA on $(\mathbb F_{p^{d_j}}^n)^\ZZ$ with matrix $M_j=M(x_1,\zeta^{(j)})$, obtained by substituting $\zeta^{(j)}$ for $x'$. By fact (d) of Section~\ref{sec:entropy},
\[
\entropy(\glorule)\geq\entropy(\glorule|_{Y_q})=\sum_{j=1}^N\entropy(H_j).
\]
Each $H_j$ is a one-dimensional GCA on the $p$-group $\mathbb F_{p^{d_j}}^n$. By Corollary~\ref{cor:gap1D}, either $\entropy(H_j)\geq\log p$, or $H_j$ is equicontinuous. In the latter case, by Lemma~\ref{lem:equichar}, $\chi_{M_j}\in\mathbb F_{p^{d_j}}[t]$. Since the determinant commutes with ring homomorphisms, $\chi_{M_j}$ is obtained from $\chi_M$ by substituting $\zeta^{(j)}$ for $x'$, and the coefficient of $x_1^{u_1}t^k$ in $\chi_{M_j}$ is $b(\zeta^{(j)})$, with $u_1\neq0$; hence $b(\zeta^{(j)})=0$. Therefore
\[
\entropy(\glorule)\geq\log p\cdot\big|\{j:\ b(\zeta^{(j)})\neq0\}\big|.
\]
Since $b$ has coefficients in $\Fp$, $b(\phi(\zeta))=b(\zeta)^p$, so the set of zeros of $b$ in $Z$ is a union of orbits, and each orbit has at most $s$ elements. Let $\tilde b=x'^{\,w}b$ be a polynomial, for a suitable $w\in\ZZ^{D-1}$, and let $\delta$ be its total degree; $\tilde b$ and $b$ have the same zeros in $Z$, and $\delta$ does not depend on $s$. By the Schwartz--Zippel lemma~\cite{Schwartz1980,Zippel1979}, applied to the grid $Z=(\mathbb E^\times)^{D-1}$, $\tilde b$ has at most $\delta q^{D-2}$ zeros in $Z$. Hence
\[
\entropy(\glorule)\geq\log p\cdot\frac{q^{D-1}-\delta q^{D-2}}{s}=\log p\cdot\frac{q^{D-2}(q-\delta)}{s},\qquad q=p^s-1.
\]
Since $D\geq2$ and $q/s\to\infty$ as $s\to\infty$, we conclude that $\entropy(\glorule)=+\infty$.
\end{proof}

\medskip\noindent
\textbf{Proof of Theorem~\ref{formula}(2)--(3) without duality.}
If all the coefficients of $\chi_M$ belong to $\Fp$, then $\glorule$ is equicontinuous by the Cayley--Hamilton argument given in the proof of Theorem~\ref{formula}(3), and $\entropy(\glorule)=0$ by Proposition~\ref{prop:equi}. Conversely, if $\glorule$ is equicontinuous, then all the coefficients of $\chi_M$ belong to $\Fp$ by Lemma~\ref{lem:equichar}. This proves (3) in every dimension. For $D\geq2$, part (2) follows from the first sentence and from Theorem~\ref{thm:noduality}. \qed

\medskip\noindent
We remark that the subsystems $Y_q$ of spatially periodic configurations make rigorous the intuition that, in dimension $D\geq2$, a non-equicontinuous GCA produces a number of independent contributions to the entropy which grows with the size of the observation window: in the proof above, these contributions are the factors $H_j$ with $b(\zeta^{(j)})\neq0$, whose number grows like $q^{D-1}/s$.

\bibliographystyle{plain}
\bibliography{bib}

\end{document}